\documentclass{elsarticle}

\usepackage{bm}
\usepackage{color}
\usepackage{graphicx}
\usepackage{graphicx}  
\usepackage{amsmath,amsfonts,amssymb}
\usepackage{hyperref}
\hypersetup{
	colorlinks=true,
	citecolor = blue,
	linkcolor=blue,
	filecolor=magenta,      
	urlcolor=cyan,
}

\usepackage{gensymb}

\newcommand\figref[1]{Fig.~\ref{#1}}

\newcommand\sectref[1]{Section~\ref{#1}}

\newcommand{\Omegarm}   {\mathrm{\Omega}}

\newcommand{\bfb}   {\mathbf{b}}
\newcommand{\bfB}   {\mathbf{B}}

\newcommand{\bfd}   {\mathbf{d}}
\newcommand{\bfD}   {\mathbf{D}}

\newcommand{\bfe}   {\mathbf{e}}
\newcommand{\bfE}   {\mathbf{E}}

\newcommand{\bfh}   {\mathbf{h}}
\newcommand{\bfH}   {\mathbf{H}}

\newcommand{\bfk}   {\mathbf{k}}

\newcommand{\bfq}   {\mathbf{q}}
\newcommand{\bfr}   {\mathbf{r}}
\newcommand{\bfs}   {\mathbf{s}}

\newcommand{\calA}  {\mathcal{A}}

\newcommand{\calM}  {\mathcal{M}}

\newcommand{\hn}   {\hat{\bf n}}

\begin{document}

\begin{frontmatter}

\title{Trefftz Approximations in Complex Media:\\
	Accuracy and Applications}  

\author{Igor Tsukerman\fnref{myfootnote}}
\address{Department of Electrical and Computer Engineering,
	The University of Akron\\
	Akron, OH 44325-3904, USA\\
	igor@uakron.edu}
\fntext[myfootnote]{Corresponding author.}

\author{Shampy Mansha, Y. D. Chong}
\address{
	School of Physical \& Mathematical Sciences,
	Nanyang Technological University\\
	21 Nanyang Link, Singapore 637371\\
	yidong@ntu.edu.sg}

\author{Vadim A. Markel}
\address{
	Department of Radiology, University of Pennsylvania, 
	Philadelphia, PA 19104, USA\\
	vmarkel@pennmedicine.upenn.edu }

%
%

\begin{abstract}
	Approximations by Trefftz functions are rapidly gaining popularity in the numerical solution
	of boundary value problems of mathematical physics. By definition, these functions 
	satisfy locally, in weak form, the underlying differential equations of the problem,
	which often results in high-order or even exponential accuracy with respect 
	to the size of the basis set. We highlight two separate examples in applied electromagnetics and photonics: 
	(i) homogenization of periodic structures,
	and (ii) numerical simulation of electromagnetic waves in slab geometries.
	Extensive numerical evidence and theoretical considerations show that
	Trefftz approximations can be applied much more broadly than is traditionally
	done: they are effective not only in physically homogeneous regions
	but also in complex inhomogeneous ones.
	Two mechanisms underlying the high accuracy of Trefftz approximations 
	in such complex cases are pointed out. The first one is related to trigonometric interpolation
	and the second one -- somewhat surprisingly -- to well-posedness of random matrices.
\end{abstract}


\end{frontmatter}




\section{Introduction}\label{sec:Intro}
%
Many classical numerical methods for partial differential equations rely on polynomial
or piecewise-polynomial approximations of the solution. Examples include traditional
finite difference (FD) schemes, the finite element method (FEM), and the boundary element method (BEM). 
But a strong incentive to achieve qualitatively higher accuracy of the numerical solution
has led, over several decades of research, to the development of Trefftz-based methods.
By definition, Trefftz functions satisfy locally (in weak form) the underlying differential equations 
of the problem, which often results in high-order algebraic or even exponential convergence 
with respect to the dimension of the basis.
This qualitative accuracy improvement has been demonstrated in a large variety of
mathematical methods and engineering applications: 
Domain Decomposition \cite{Herrera00,Farhat-DD-DG09},
Generalized FEM \cite{Melenk96,Babuska97,Babuska-GFEM2004,Plaks03,Proekt02,
Strouboulis-GFEM-Helmholtz2006}, 
Discontinuous Galerkin \cite{Farhat-DD-DG09,Cockburn00,Arnold02,Buffa-Monk-ultraweak08,Gittelson09,
Gabard-wave-based-DG-UW-LS11,Hiptmair-PWDG-Helmholtz2011,Kretzschmar-IMA16},
and finite difference (``Flexible Local Approximation MEthods,'' FLAME) \cite{Tsukerman-JCP10,Tsukerman05,Tsukerman06,Tsukerman-PBG08}.

It is not our intention to review all, or even some, of these Trefftz-oriented
methods; several good reviews are already available:
\cite{Deckers-wave-method-overview2014,Qin-Trefftz-FEM2005} and
especially \cite{Hiptmair2016}. Rather, our focus is on one question 
central in these methods: \textit{why are Trefftz approximations so effective}?

\begin{figure}\label{fig:test-waves}
	\centering
	\includegraphics[width=0.4\linewidth]{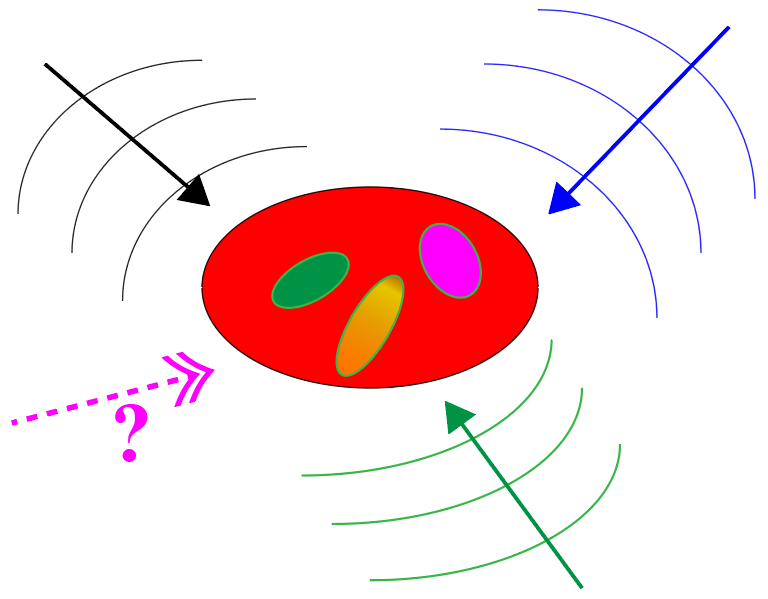} \hskip 0.2in
	\includegraphics[width=0.4\linewidth]{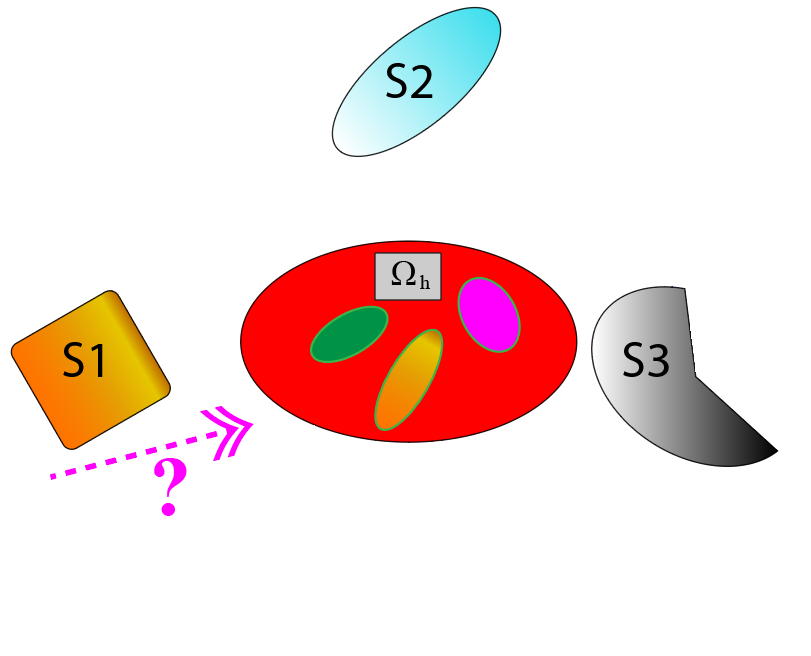}
	\caption{Several incident waves (schematically indicated with solid arrows) 
		give rise to the respective total fields
		inside an inhomogeneous scatterer. These total fields, by definition, 
		form a Trefftz set. This ``training set'' can be precomputed and then used to approximate the field 
		induced by another wave, indicated with a dashed arrow and a question mark. 
		For visual clarity, only three Trefftz waves are sketched, and only the incident components.
		Left: one inhomogeneous object is present. Right: additional scatterers 
		(such as S1, S2, S3) may be present in the case of an unknown
		field (dashed arrow) whose approximation is sought within a given small subdomain $\Omega_h$.}
\end{figure}

A simplified intuitive picture is shown in \figref{fig:test-waves}, left panel.
Several incident waves, schematically indicated with solid arrows, are
impinging on an object (in general, physically inhomogeneous) and give rise to the respective 
total fields inside and to scattered fields outside that object. 
For visual clarity, only the incident waves
are sketched in the figure, and their number is limited to three.

The total fields inside the scatterer, by definition, form a Trefftz set.
One may view it as a ``database'' or ``training set,''
which can be precomputed and then used to approximate the field induced by another wave,
indicated with a dashed arrow and a question mark in \figref{fig:test-waves}.
This approximation of one physically meaningful solution by other
physically meaningful solutions (as opposed to, say, generic polynomials)
certainly makes intuitive sense but is not trivial from the mathematical perspective.

The right panel of \figref{fig:test-waves} illustrates a more interesting, and more complicated,
case. Suppose that the Trefftz training set has been generated for the original inhomogeneous 
scatterer -- same as in the left panel. However, the unknown ``dashed arrow'' solution 
\textit{may involve additional objects} -- such as S1, S2, S3 -- in the computational domain.
Obviously, under this complication, little can be inferred about the unknown solution
from the Trefftz set \textit{in the whole domain}, especially in the regions around the
additional scatterers. One may hope, however, that the field \textit{within a given small subdomain} 
$\Omega_h$ \textit{inside the original scatterer} can still be  approximated accurately
as a superposition of the known Trefftz waves. This setup is the central issue of 
Sections~\ref{sec:Auxiliary-basis}, \ref{sec:Random-matrices}, and is inspired by our numerical
experiments with pseudorandom structures of \sectref{sec:FLAME-slab}, as well as by our
earlier work on multiparticle problems \cite{Dai-Webb11}.

The overall motivation for the paper is to highlight applications of Trefftz functions to problems
involving complex, inhomogeneous media. Much of mathematical analysis so far
has revolved around the homogeneous case (that is, equations with constant coefficients),
where cylindrical, spherical or plane waves serve as Trefftz functions for the Helmholtz equation, 
while harmonic polynomials are used for the Laplace equation. 
One can refer, for example, to papers by Melenk, Hiptmair, Moiola, Perugia 
\textit{et al.} cited above, to the references in these papers, 
and to Perrey-Debain's paper \cite{Perrey-Debain-PW-convergence2006}.
Much less attention has been paid to the inhomogeneous case 
\cite[Chapter IV]{Melenk95}, \cite[Section 3]{Melenk99},
\cite{Laghrouche-jumps2005,Imbert-Gerard-interp-PW2015},
which is substantially more complicated but at the same time more rewarding in practice.

For illustration, in Sections \ref{sec:Trefftz-homogenization} and \ref{sec:FLAME-slab}
we consider two application examples where Trefftz approximations prove to
be effective for two different variations of the generic setup shown 
in \figref{fig:test-waves}. The first example is non-asymptotic and nonlocal 
two-scale homogenization. Instead of a single
scatterer, in this case one deals with a periodic structure;
Trefftz functions on the fine scale are Bloch waves traveling in different directions,
and on the coarse scale -- the corresponding plane waves.

The second example involves a common setup in metasurface and nanophotonics research:
a patterned finite-thickness slab. This problem is especially challenging computationally
when the pattern is non-periodic and the slab is geometrically large relative
to the vacuum wavelength. One possible simulation procedure relies on high-order 
Trefftz difference schemes (FLAME). The Trefftz bases are computed ``locally,''
i.e. over relatively small segments of the structure (\sectref{sec:FLAME-slab}).

Sections~\ref{sec:Trig-interpolation} and \ref{sec:FD-Trefftz} provide
background information needed in the application examples of
Sections~\ref{sec:Trefftz-homogenization} and \ref{sec:FLAME-slab}.
The underlying mechanisms for the accuracy of Trefftz approximations
are discussed in Section~\ref{sec:Trefftz-approximation}.

\section{Preliminaries: Trigonometric Projection and Interpolation}\label{sec:Trig-interpolation}
%
Trigonometric approximation of periodic functions is a well-established subject.
Here we summarize the key mathematical results that will be needed in Section~\ref{sec:Interpolation-argument}.

For any Lipschitz-continuous periodic function $g$ on $[-\pi, \pi]$,
one may consider its best possible approximation by a trigonometric polynomial $T_n$ in the maximum norm:
\begin{equation}\label{eqn:En-best-approx-trig-poly}
E_n^T(g) \,=\, \min_{\alpha, \beta} \max_{\phi \in [-\pi, \pi]}
\left| g(\phi) - T_n(\alpha, \beta, \phi) \right|, ~~~
\end{equation}
where
\begin{equation}\label{eqn:trig-poly-defined}
T_n(\alpha, \beta, \phi) \, \equiv \, \alpha_0 + \sum_{\nu=1}^n 
(\alpha_\nu \cos \nu \phi + \beta_\nu \sin \nu \phi),
\end{equation}
$$
\alpha \equiv \{\alpha_0, \alpha_1, \ldots, \alpha_n\},
~~~ \beta \equiv \{\beta_1, \ldots, \beta_n\}
$$
A slightly modified notation of \cite{Meinardus-Approximation-theory1967} is used here.
Note that the total number of coefficients $\alpha$, $\beta$ in the trigonometric series
is $N = 2n+1$.

It follows from Jackson's theorem \cite{Jackson-Theory-approximation1930}, 
or \cite[Theorem 41]{Meinardus-Approximation-theory1967},
that if the derivative $g^{(l+1)} (\phi)$ exists and is bounded, i.e.
\begin{equation}\label{g-bounded-derivative}
\left| g^{(l+1)} (\phi) \right|  \, \leq \, M_{l+1},
~~~ l = 0,1, \ldots
\end{equation}
then
\begin{equation}\label{Eg-leq-nk1}
E_n^T(g) \, \leq\, \frac{c^{l+1} M_{l+1}} {n^{l+1}}, ~~~
c = 1 + \frac{\pi^2}{2}
\end{equation}
For reasons that will become apparent in Section~\ref{sec:Interpolation-argument}, 
we are interested primarily
in trigonometric \textit{interpolation} rather than the best approximation,
and thus need to relate the two. The interpolant $\tilde{T}_N(\phi)$ 
of a given function $g(\phi)$ over a set of $N = 2n + 1$ equidistant knots
$\{\phi_m \}$ is defined in a standard way, by requiring that
\begin{equation}\label{eqn:interpolant-defined}
   \tilde{T}_N(g, \phi_m) = g(\phi_m), ~~ \phi_m = \frac{2\pi m}{N},
   ~~~m = 0,1, \ldots, N-1
\end{equation}
It is known that this interpolant exists and is unique. Furthermore,
there is an upper bound for the interpolation error:
\begin{equation}\label{eqn:interp-error-vs-approximation-error}
   \| g - \tilde{T}_N(g) \|_\infty ~ \leq ~ (1 + \Lambda_N) \,
   \| g - T_N \|_\infty ~\equiv~ (1 + \Lambda_N) \, E_n^T(g)
\end{equation}
where $\Lambda_N$ is the Lebesgue constant,
which itself has an upper bound \cite{Cheney1975}
\begin{equation}\label{eqn:Lebesgue-const-upper-bound}
\Lambda_N \, \leq \, 2 \pi^{-1} \log N + \frac{5}{3}
\end{equation}
All the above information can be found in a variety of sources,
including very recent ones \cite{Austin-Trefethen-trig-interp2017,Austin-PhD-thesis-interpolation2016},
\cite[Section 7]{Trefethen-Weideman-exp-convergent-trapezoidal2014}.

Combining \eqref{eqn:interp-error-vs-approximation-error}, \eqref{eqn:Lebesgue-const-upper-bound},
and \eqref{Eg-leq-nk1}, one has
\begin{equation}\label{eqn:interp-error-algebraic-convergence-vs-N}
\| g - \tilde{T}_N(g) \| \, \leq \, \left(2 \pi^{-1} \log N + \frac{8}{3} \right) 
\left(1 + \frac{\pi^2}{2} \right)^{l+1} \frac{M_{l+1}}  {n^{l+1}}
\end{equation}
This indicates fast uniform algebraic convergence of the interpolant with respect
to the number of knots. Moreover, under additional assumptions of analyticity
of $g(\theta)$ in a strip of the complex plane 
Re$\, \theta \in (0, 2\pi)$, $|\mathrm{Im} \, \theta | < \delta$,
convergence becomes exponential \cite[(7.19)]{Trefethen-Weideman-exp-convergent-trapezoidal2014}:
\begin{equation}\label{eqn:interp-error-exp-convergence-vs-N}
\| g - \tilde{T}_N(g) \| \, \leq \, \frac{4M \exp[-\delta(N+1)/2]}
{1 - \exp(-\delta)}
\end{equation}
We are also interested in the approximation of the integral
\begin{equation}\label{eqn:integral-of-g}
I \,=\, \int_0^{2\pi} g(\theta) \, d \theta
\end{equation}
using the values of $g$ at the equispaced knots:
\begin{equation}\label{eqn:num-quadrature-of-g}
I_N \,=\, \frac{2\pi}{N} \, \sum_{m=1}^{N-1} g(\theta_m),
~~~ \theta_m = \frac{2\pi m}{N},
\end{equation}
which is the trapezoidal rule for the numerical quadrature.
Under the same analyticity assumptions as above, the error of this
quadrature can be bounded as \cite[(7.20)]{Trefethen-Weideman-exp-convergent-trapezoidal2014}
\begin{equation}\label{eqn:quadrature-error-exp-convergence-vs-N}
\| I_N - I \| \, \leq \, \frac{8\pi M \exp[-\delta(N+1)/2]}
{1 - \exp(-\delta)}
\end{equation}
A similar result can be found in \cite[Theorem 1]{Austin-Trefethen-trig-interp2017}.
Adapted to our needs and notation, it states:

\vskip 0.1in
\noindent
If $f$ is $l$ times continuously differentiable and $f^{(l)}$ is Lipschitz continuous, 
then
\begin{equation}\label{eqn:Austin-Trefethen-trig-interp-accuracy-power}
| I - \tilde I_N^{} | ,  \| f - \tilde t_N^{} \| 
= \mathcal{O}(N^{-(l+1)}).	
\end{equation}
If $f$ can be analytically continued to a $2\pi$-periodic
function for $-\delta < \mathrm{Im} \, x < \delta$ for some $\delta > 0$, then
for any $\hat \delta < \delta$,
\begin{equation}\label{eqn:Austin-Trefethen-trig-interp-accuracy-exp}
| I - \tilde I_N^{} | , \| f - \tilde t_N^{} \| = 
\mathcal{O}(\exp(-\hat \delta N)).
\end{equation}
\vskip 0.1in

The qualitative conclusion of this section is that \textit{trigonometric interpolation
of a smooth periodic function provides a very accurate approximation of the function
and its integrals.}

\section{Preliminaries: Finite Difference Trefftz Schemes}\label{sec:FD-Trefftz}
%
Another preliminary subject, which will be needed in \sectref{sec:FLAME-slab}, is FLAME
\cite{Tsukerman-JCP10,Tsukerman05,Tsukerman06,Tsukerman-PBG08,Tsukerman-book07,Pinheiro07}.
Recall that classical FD schemes are typically derived from Taylor expansions; but
this is problematic if the solution is not sufficiently smooth -- e.g. at material interfaces.
That is the root cause of the notorious ``staircase'' effect at slanted or curved interface boundaries
that do not conform geometrically to the grid lines.
FLAME replaces Taylor polynomials with Trefftz functions, which often produces
high-order schemes.

The key ideas of FLAME are as follows.
Let a boundary value problem be defined in a computational domain $\Omega$
and consider a small subdomain $\Omega_h$ within which a difference scheme 
is to be formed. In $\Omega_h$, introduce a set of $m$ degrees of freedom (DoF). 
These DoF are, by definition, linear functionals, $l_\beta(u)$
($\beta = 1,2,\dots, m$), each mapping any admissible field $u$ to a number
(real or complex, depending on the problem). The simplest example of DoF for a scalar field
$u$ is as set of nodal values $l_\beta(u) \equiv u(\mathbf{r}_\beta)$, 
where $\mathbf{r}_1,\dots,\mathbf{r}_m$ are a
set of grid nodes in $\Omega_h$. In the case of vector fields, 
one may also consider fluxes, circulations, etc. as other examples of DoF .

Locally, within $\Omega_h$, the solution $u$ is approximated by a
linear combination of Trefftz functions $\psi_\alpha$ ($\alpha = 1,2,\dots, n$) :
\begin{equation}\label{eqn:uh-eq-c-psi}
   u(\mathbf{r}) \,\approx\, u_h(\mathbf{r}) \equiv 
   \sum\nolimits_{\alpha=1}^n c_\alpha \psi_\alpha (\mathbf{r}) \,=\,
   \underline{c}^T \underline{\psi}(\mathbf{r}),
\end{equation}
where $\underline{c} \in \mathbb{C}^n$ is a coefficient vector and
$\underline{\psi}$ is a vector of basis functions (both generally
complex). In $\Omega_h$, we seek an FD equation of the form
\begin{equation}\label{eqn:s-beta-l-beta-eq-0}
    \sum\nolimits_{\beta=1}^m s_\beta l_\beta (u) = 0,
\end{equation}
where $\underline{s} = (s_1, s_2,\dots, s_m)^T$ is a vector of complex
coefficients (a ``scheme'') to be determined.  In the simplest version
of FLAME, the scheme is required to be exact for any linear
combination \eqref{eqn:uh-eq-c-psi} of basis functions.
Then, after straightforward algebra, one obtains \cite{Tsukerman05,Tsukerman06}
\begin{equation}\label{eqn:s-in-null-Nt}
\underline{s} \in \mathrm{Null}(N^T), ~~~
\mathrm{where}~~  N^T_{\alpha \beta} = l_\beta (\psi_\alpha).
\end{equation}
There are also least-squares versions of this idea \cite{Boag94,Tsukerman-JCP10}.

Many illustrative examples are given in \cite{Tsukerman-book07,Tsukerman05,Tsukerman06}.
Here we mention just one of them, closely related to the construction of
FLAME schemes in \sectref{sec:FLAME-slab}. For the 2D Helmholtz equation, one may
consider a Trefftz basis set of eight plane waves
traveling at the angles $\phi_0 + m \pi/4$ ($m = 0,1, \ldots, 7$), where $\phi_0$ is a given angle;
practical choices are $\phi_0 = 0$ or $\phi_0 = \pi / 8$.
Evaluating these plane waves over a standard $3 \times 3$ grid ``molecule,'' 
one obtains an $8 \times 9$ matrix $N^T$  whose null vector is the FLAME scheme. 
The result for $\phi_0 = 0$ is a nine-point ($3 \times 3$) order-six scheme
\cite{Tsukerman06}. For $\phi_0 = \pi/8$, one arrives at a scheme derived 
by Babu\v{s}ka \textit{et al}. in 1995 \cite{Babuska-Ihlenburg95} from very different considerations.

%
\section{Trefftz Homogenization of Electromagnetic Structures}
\label{sec:Trefftz-homogenization}
%
We consider Trefftz-based homogenization 
of electromagnetic periodic structures (photonic crystals and metamaterials).
The general description of the problem in this section follows
\cite{Tsukerman-PLA17,Tsukerman-Markel14} closely; but our focus here
is on \textit{Trefftz approximation}, the importance of other aspects of the problem notwithstanding.

The physical essence of the problem is as follows. A sample of a periodic material
is illuminated by incoming monochromatic electromagnetic waves
at a given frequency $\omega$ and the corresponding free-space wavenumber
$k_0 = \omega / c$. To sidestep the complicated problem of field behavior
at corners, the sample is assumed to be a finite-thickness slab contained
between the planes $z = 0$ and $z = L$, and infinite in the $x$ and $y$ directions. 
The periodic medium in the sample is to be replaced with a homogeneous material in such a way 
that the scattering wave pattern would be preserved as accurately as possible.

Following \cite{Tsukerman-PLA17,Tsukerman-Markel14}, let us define the problem
more precisely. Assume that the intrinsic dielectric permittivity $\tilde{\epsilon}(\bfr)$ 
within the slab is lattice-periodic, and that all material constituents are nonmagnetic,
$\tilde{\mu}({\bf r}) = 1$. Let all constitutive relationships be local and linear,
and let the sample be illuminated by monochromatic waves with a given far-field pattern; 
these waves are reflected by the metamaterial.

The problem has two principal scales (levels). 
\textit{Fine-level} fields are the exact solutions of Maxwell's equations 
for given illumination conditions for a given sample. 
These fields are denoted with small letters $\bfe$, $\bfd$,
$\bfh$ and $\bfb$. In general, their variation in space is rapid and
consistent with the microstructure of metamaterial cells.
\textit{Coarse-level} fields  $\bfE$, $\bfD$, $\bfH$, $\bfB$
vary on a characteristic scale greater that the cell size. 
They represent some smoothed (averaged) versions of the fine-level fields and are
auxiliary mathematical constructions rather than measurable physical quantities.
The coarse-level fields are sought to satisfy Maxwell's equations 
\textit{and all interface boundary conditions} as accurately
as possible. 

Importantly, effective magnetic properties of metamaterials cannot be determined 
from the bulk behavior alone \textit{as a matter of principle}. This is due, in particular,
to the fact that the Maxwell equation $\nabla \times \bfH = -i k_0 \bfD$  is invariant with respect to 
an arbitrary simultaneous rescaling of vectors \textbf{H} and \textbf{D}. 
Loosely speaking, bulk behavior defines the dispersion relation only, 
while magnetic characteristics depend on the boundary impedance as well.

The fine-level fields satisfy macroscopic Maxwell's equations of the form
\begin{equation}\label{eq:Maxwell_exact}
   \nabla \times \bfh(\bfr) = -i k_0 \tilde{\varepsilon}(\bfr) \, \bfe(\bfr) \ , \ \ ~~
   \nabla \times \bfe (\bfr) = i k_0 \, \bfh(\bfr)
\end{equation}
\noindent
everywhere in space, supplemented by the usual radiation boundary
conditions at infinity. Outside the slab, the most general solution
of \eqref{eq:Maxwell_exact} can be written as a superposition of
incident, transmitted and reflected waves. For the electric field, we
can write these in the form of angular-spectrum expansions \cite{Tsukerman-Markel14}:
\begin{subequations}
\label{eq:Efine}
\begin{eqnarray}
\label{eq:Ei}
&& \bfe_i(\bfr) = \int \bfs_i(k_x, k_y) e^{i\left(k_x x + k_y y +
	k_z z \right) } dk_x dk_y \ ,    \\
\label{eq:Et}
&& \bfe_t(\bfr) = \int \bfs_t(k_x, k_y)
e^{i\left(k_x x + k_y y + k_z z \right)} d k_x d k_y \ , \ \ z > L \ ,    \\
\label{eq:Er}
&& \bfe_r(\bfr) = \int \bfs_r(k_x, k_y) e^{i \left(k_x x + k_y y - k_z
	z \right)} d k_x d k_y \ , \ \ z < 0 \ ,
\end{eqnarray}
\end{subequations}
\noindent
where
\begin{equation}\label{eq:disp_vac}
   k_z = \sqrt{k_0^2 - k_x^2 - k_y^2} \ ,
\end{equation}
\noindent
and the square root branch is defined by the condition $0 \leq \arg k_z < \pi$. 
Expressions for the magnetic field are obtained
from \eqref{eq:Efine} by using the second Maxwell equation in
\eqref{eq:Maxwell_exact}. In \eqref{eq:Efine}, $\bfs_i(k_x,k_y)$,
$\bfs_t(k_x,k_y)$ and $\bfs_r(k_x,k_y)$ are the angular spectra of the
incident, transmitted and reflected fields. Waves included in these
expansions can be evanescent or propagating. For propagating
waves, $k_x^2 + k_y^2 < k_0^2$, otherwise the waves are evanescent.

Everywhere in space, the total electric field $\bfe(\bfr)$ can be
written as a superposition of the incident and scattered fields, viz,
\begin{equation}
\label{eqn:ei_es}
\bfe(\bfr) = \bfe_i(\bfr) + \bfe_s(\bfr) \ .
\end{equation}
\noindent
Outside the material, the reflected and transmitted fields form the
scattered field:
\begin{equation}
\label{eqn:es-eq-er-et}
\bfe_s(\bfr) = \left\{
\begin{array}{ll}
\bfe_r(\bfr) \ , & z < 0 \ , \\
\bfe_t(\bfr) \ , & z > L \ .
\end{array} \right.
\end{equation}
\noindent
The scattered field \emph{inside} the material is also formally
defined by \eqref{eqn:ei_es}.

It is natural to approximate fine-level fields via a basis set of Bloch waves 
traveling in different directions:
\begin{equation}\label{eqn:eh-Bloch}
\bfe_{m \alpha}(\bfr) = \tilde{\bfe}_{m \alpha}(\bfr) \exp(i \bfq_{m \alpha} \cdot \bfr) \ , \ \
\bfh_{m \alpha} = \tilde{\bfh}_{m \alpha}(\bfr) \exp(i \bfq_{m \alpha}
\cdot \bfr) \ ,
\end{equation}
where index $\alpha$ labels both the wave vector and the polarization
state of the Bloch wave in a lattice cell $m$;
$\tilde{\bfe}_{m \alpha}({\bf r})$, $\tilde{\bfh}_{m \alpha}({\bf r})$ are
the respective lattice-periodic factors. 
As the notation indicates, the basis is defined cell-wise; different bases in 
different lattice cells could be used. This makes the homogenization problem tractable 
and reducible to a single cell, rather than global and encompassing the whole sample.

On the coarse scale, a natural counterpart of the fine-scale Bloch basis
is a set of generalized plane waves
\begin{equation}\label{eqn:Psi-eq-E0-H0-exp}
\Psi_{m \alpha} = \{ \mathbf{E}_{m \alpha}, \mathbf{H}_{m \alpha} \} = 
\{ \mathbf{E}_{0 m \alpha}, \mathbf{H}_{0m \alpha} \} \exp(i \bfq_{m \alpha} \cdot \bfr)
\end{equation}
which satisfy Maxwell's equations in a homogeneous but possibly anisotropic medium;
subscript `0' indicates the field amplitudes to be determined.

Further technical details of the procedure can be found in \cite{Tsukerman-Markel14,Tsukerman-PLA17}.
The final result is as follows. First, the coarse-level wave vector
for each plane wave is taken to be the same as its counterpart for the corresponding Bloch wave,
which is already reflected in our notation above \eqref{eqn:eh-Bloch}, \eqref{eqn:Psi-eq-E0-H0-exp}. 
Secondly, the amplitudes $\{ \mathbf{E}_{0 m \alpha}, \mathbf{H}_{0m \alpha} \}$  
of each plane wave are the \textit{boundary} average
of the \textit{tangential} components of the respective fine-scale Bloch wave:
\begin{equation}\label{eqn:EH0-eq-face-avrg-eh}
\bfE_{0m \alpha} \,=\, \calA_{m} ^{\, \tau} \tilde{\bfe}_{m \alpha}, ~~~
\bfH_{0m \alpha} \,=\, \calA_{m}^{\, \tau} \tilde{\bfh}_{m \alpha}
\end{equation}
The averaging operator $\calA_{m}^{\, \tau}$ for tangential components of a generic
vector field $\mathbf{f}$
is defined, in the case of an orthorhombic cell $\mathbb{C}_{m}$, as
\begin{equation}\label{eqn:bdry-avrging-operator-tau}
(\calA_m^{\, \tau})_{\gamma} \, \mathbf{f} \,\equiv\, 
\frac{
\int_{\partial \mathbb{C}_{m}} f_{\gamma} \, |\hn \times \hat{\bfr}_{\gamma}| \, dS
}
{
\int_{\partial \mathbb{C}_{m}} |\hn \times \hat{\bfr}_{\gamma}| \, dS
}
, ~~~ \gamma = 1,2,3;  ~~ \hat{\bfr}_{1,2,3} = \hat{x}, \hat{y}, \hat{z}
\end{equation}
%
Here $|\hn \times \hat{\bfr}_{\gamma}|$ 
acts simply as the Kronecker delta for the faces of the cell parallel 
to a given coordinate direction $\hat{\bfr}_{\gamma}$, $\gamma = 1,2,3$.
Note that the averages in \eqref{eqn:EH0-eq-face-avrg-eh} 
involve the \textit{periodic factor} of the Bloch wave.
The amplitudes $\bfE_{0m \alpha}$, $\bfH_{0m \alpha}$,
along with the Bloch wave vector, define the coarse-level basis function $\alpha$ in a lattice cell $m$.

The homogenization procedure of \cite{Tsukerman-PLA17,Tsukerman-Markel14} 
leads to a system of algebraic equations of the form
\begin{equation}\label{eqn:Psi-DB-eq-M-Psi-EH}
\Psi_{DB} \overset{l.s.}{=} \calM \Psi_{EH}
\end{equation}
Here `l.s.' stands for `least squares'.
Each column of the rectangular matrix $\Psi_{EH}$
corresponds to a given coarse-level basis function $\alpha$, and the entries
of that column are the $xyz$-components of the wave amplitudes $\bfE_{0m \alpha}$, $\bfH_{0m \alpha}$.
The number of columns $n$ is equal to the chosen number of basis functions;
the number of rows is, in general, six, unless some of the field components
are known to be zero (e.g. for $s$- or $p$-polarized waves).
The $\Psi_{DB}$  matrix is completely analogous 
and contains the $\mathbf{DB}$ amplitudes derived from Maxwell's curl equations:
\begin{equation}\label{B0-D0-eq-q-times-EH}
\bfB_{0m \alpha} = k_0^{-1} \bfq_{m \alpha} \times \bfE_{0m \alpha}, ~~
\bfD_{0m \alpha} = -k_0^{-1} \bfq_{m \alpha} \times \bfH_{0m \alpha}
\end{equation}
The (local) material tensor is represented, in general, by a $6 \times 6$ matrix. 
Since the number of columns in matrix $\Psi_{EH}$ is typically greater than the number of rows,
the matrix equation \eqref{eqn:Psi-DB-eq-M-Psi-EH} for the material tensor is solved in the least squares sense:
\begin{equation}\label{eqn:M-eq-DB-over-EH}
\calM \,=\, \Psi_{DB} \Psi_{EH}^+;    
~~~~~ \delta_{\mathrm{l.s.}} \,=\, \| \Psi_{DB} - \calM \Psi_{EH} \|_2
\end{equation}
where $\Psi_{EH}^+$ is the Moore-Penrose pseudoinverse of $\Psi_{EH}$, and
$\delta_{\mathrm{l.s.}}$ is the associated least-squares error.

\begin{figure}
\centering
\includegraphics[width=0.85\linewidth]{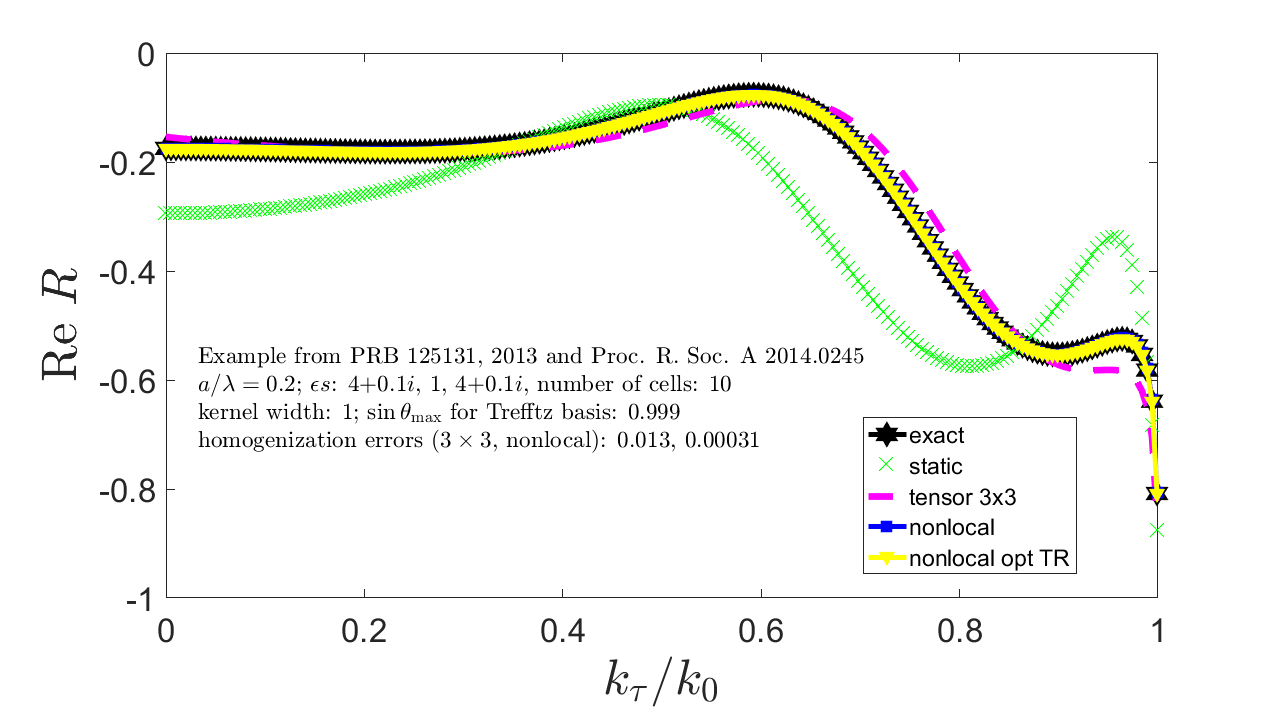}
\includegraphics[width=0.85\linewidth]{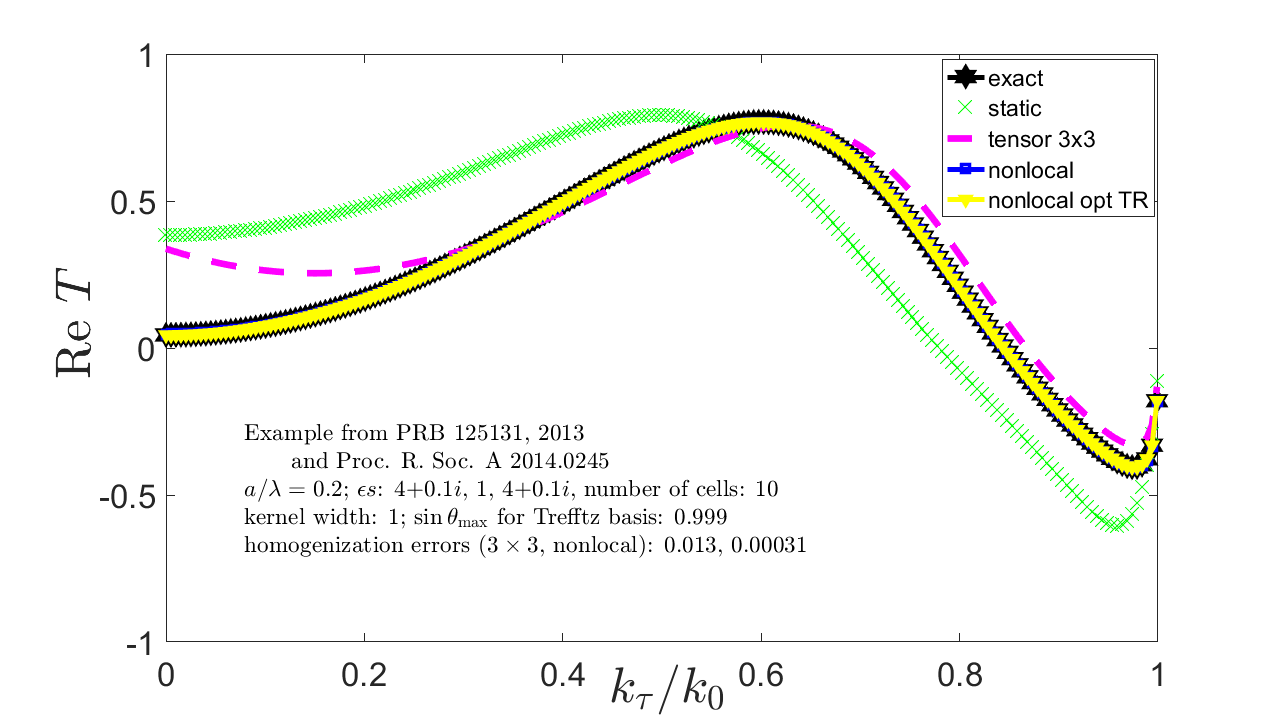}
\caption{Example A of a layered medium from \cite{Markel-Tsukerman-PRB2013,Tsukerman-Markel14}.
	The real part of $R$ (left) and $T$ (right) vs. the sine
	of the angle of incidence; non-asymptotic and nonlocal homogenization.
	The lattice cell contains three layers of widths $a/4$, $a/2$ and $a/4$,
	with scalar permittivities $\epsilon_1$, $\epsilon_2$, and $\epsilon_1$, respectively.
	($\epsilon_1 = 4 + 0.1i$ and $\epsilon_2 = 1$.) Fine-level basis: 
	$2n_{\mathrm{dir}}$ Bloch modes traveling at $n_{\mathrm{dir}} = 7$ different angles 
	in $(-\pi/2, \pi/2)$; $n_{\mathrm{dir}} = 7.$
	The kernel width parameter $\tau_0 = a$.
	The reflection and transmission coefficients from nonlocal homogenization
	are visually indistinguishable from the exact ones. The nonlocal procedure
	includes two additional DoF: the convolution integrals of the tangential
	components of the electric and magnetic fields.}
\label{fig:Re-RT-vs-angle-ExA-PRB2013}
\end{figure}

\begin{figure}
\centering
\includegraphics[width=0.85\linewidth]{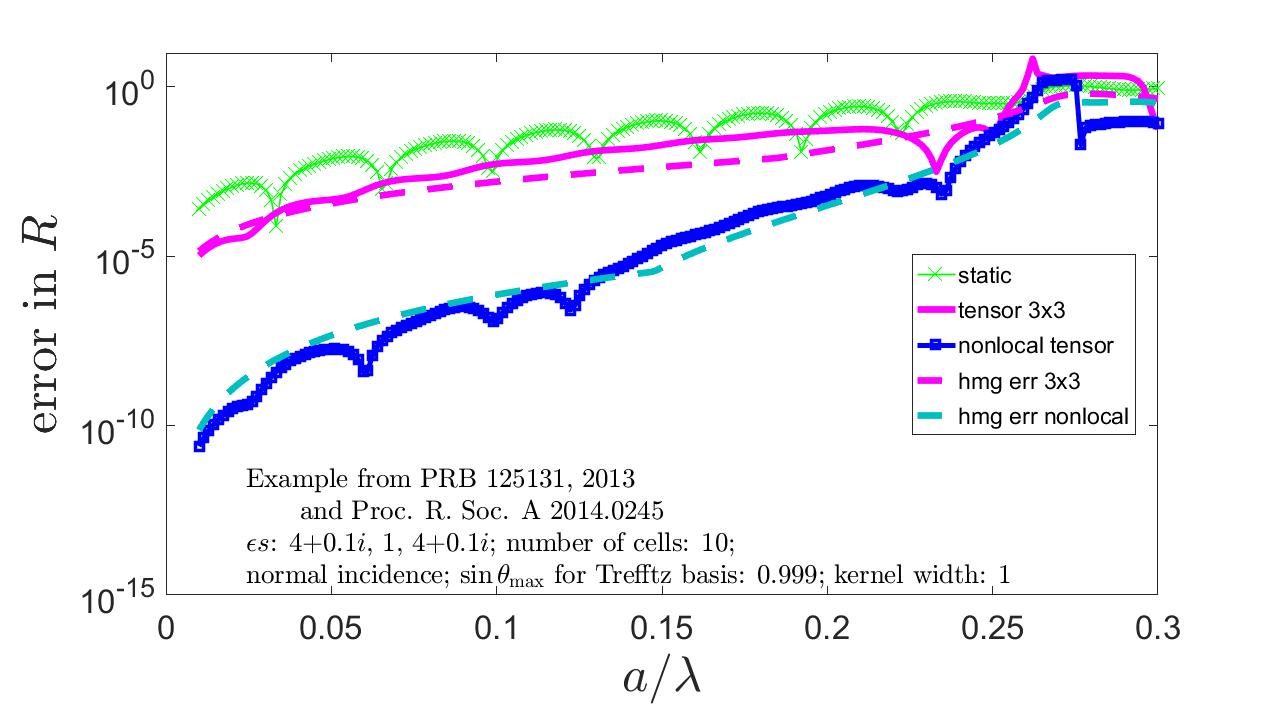}
\includegraphics[width=0.85\linewidth]{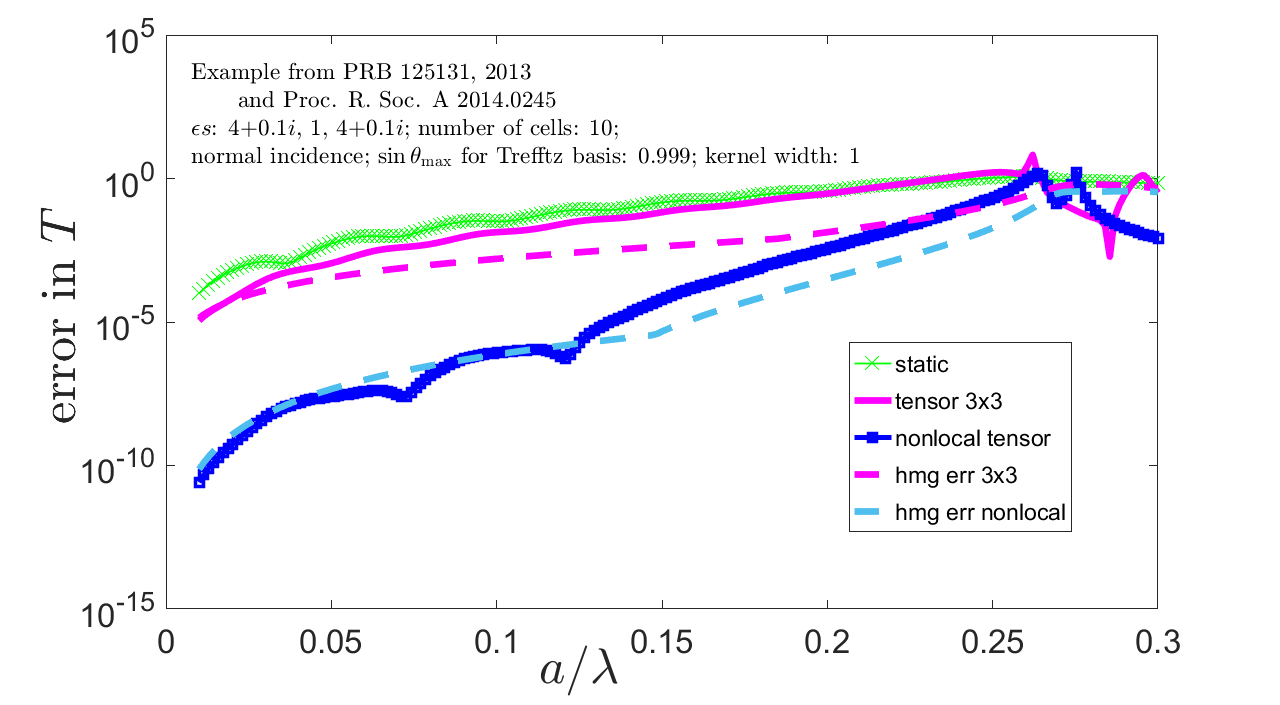}
\caption{Example A of a layered medium from \cite{Markel-Tsukerman-PRB2013,Tsukerman-Markel14}.
	Absolute error in $R$ (left) and $T$ (right) vs. $a / \lambda$; 
	non-asymptotic and nonlocal homogenization.
	The lattice cell contains three layers of widths $a/4$, $a/2$ and $a/4$,
	with scalar permittivities $\epsilon_1$, $\epsilon_2$, and $\epsilon_1$, respectively.
	($\epsilon_1 = 4 + 0.1i$ and $\epsilon_2 = 1$.) Fine-level basis: 
	$2n_{\mathrm{dir}}$ Bloch modes	traveling at $n_{\mathrm{dir}} = 7$ 
	different angles in $(-\pi/2, \pi/2)$; $n_{\mathrm{dir}} = 7.$
	The kernel width parameter $\tau_0 = a$.
	The accuracy of the nonlocal procedure is, by far, the highest.
	The nonlocal procedure
	includes two additional DoF: the convolution integrals of the tangential
	components of the electric and magnetic fields.
}
\label{fig:error-R-vs-angle-ExA-PRB2013}
\end{figure}

As demonstrated in \cite{Tsukerman-PLA17}, the homogenization accuracy can be
further improved by including, in addition to the $EH$ amplitudes,
integral DoF of the form
\begin{equation}\label{eqn:D-eq-Eps-star-E}
\bfD(\bfr) \,=\, \int_{\Omegarm} \mathcal{E}(\bfr, \bfr') \, \bfE(\bfr') \, d \Omegarm
\end{equation}
where $\mathcal{E}$ is a convolution kernel depending only on 
the coordinates tangential to the boundary of the sample:
$$
\mathcal{E}(\bfr, \bfr') = 
\mathcal{E}(\hn \times \bfr, \hn \times \bfr')
$$
A natural (but certainly not unique) choice for this kernel is a Gaussian 
$$
\mathcal{E}(\bfr, \bfr') = 
\mathcal{E}_0 \exp(-\tau_0^{-2} \, |\hn \times (\bfr - \bfr')|^2 )
$$
where the amplitude $\mathcal{E}_0$ and width $\tau_0$ are adjustable parameters, 
and $\hn$ is the unit normal vector. 

Since our focus is on the approximation properties of Trefftz functions
and not on the homogenization procedure per se, we do not discuss the physics
of the problem here, or the merits and demerits of nonlocal vs. local theory.\footnote{
It should, however, be noted that our nonlocal procedure operates in real space,
in contrast with $k$-space techniques that we critiqued elsewhere \cite{Markel-Tsukerman-PRB2013}.}
We also omit further technical details
and limit ourselves to just one illustration example.

Shown in Figs.~\ref{fig:Re-RT-vs-angle-ExA-PRB2013} and \ref{fig:error-R-vs-angle-ExA-PRB2013}
are the reflection $R$ and transmission $T$ coefficients for electromagnetic waves
propagating through a layered slab. These coefficients are defined in a standard
way, as the ratio of the complex amplitudes of the reflected/transmitted waves
to that of the incident wave. The geometric and physical parameters correspond
to Example A of \cite{Markel-Tsukerman-PRB2013}: the lattice cell of a width $a$
contains three layers of widths $a/4$, $a/2$ and $a/4$,
with scalar permittivities $\epsilon_1$, $\epsilon_2$, and $\epsilon_1$, respectively;
$\epsilon_1 = 4 + 0.1i$ and $\epsilon_2 = 1$. The fine-level Trefftz basis
contains $2n_{\mathrm{dir}}$ Bloch modes traveling at $n_{\mathrm{dir}} = 7$ equispaced angles 
in $(-\pi/2, \pi/2)$; $n_{\mathrm{dir}} = 7$. In nonlocal homogenization,
the additional DoF are the integrals of the form \eqref{eqn:D-eq-Eps-star-E},
with the Gaussian kernel of width $\tau_0 = a$.

Fig.~\ref{fig:Re-RT-vs-angle-ExA-PRB2013} shows the real part of $R$ and $T$
as a function of the angle of incidence, for $a / \lambda = 0.2$.
(The imaginary parts are not plotted to save space but are qualitatively similar).
Since analytical solutions for wave propagation in layered media are fairly
simple and well known, one may easily calculate the errors in $R$ and $T$;
those are plotted in Fig.~\ref{fig:error-R-vs-angle-ExA-PRB2013}.

The figures show that our numerical results, especially for nonlocal homogenization,
are highly accurate. In fact, we are not aware of any alternative methods
that could produce a comparable level of accuracy at a comparable computational cost.\footnote
{The latter provision is needed to exclude from consideration ``brute force''
numerical optimization of the material tensor.}

What explains this high accuracy? Plausible mechanisms are presented 
in \sectref{sec:Trefftz-approximation}.

\section{Electromagnetic Waves in Slab Geometries}\label{sec:FLAME-slab}
%
\subsection{Formulation of the problem}\label{sec:Formulation}
%
\begin{figure}
	\centering
	\includegraphics[width=11cm]{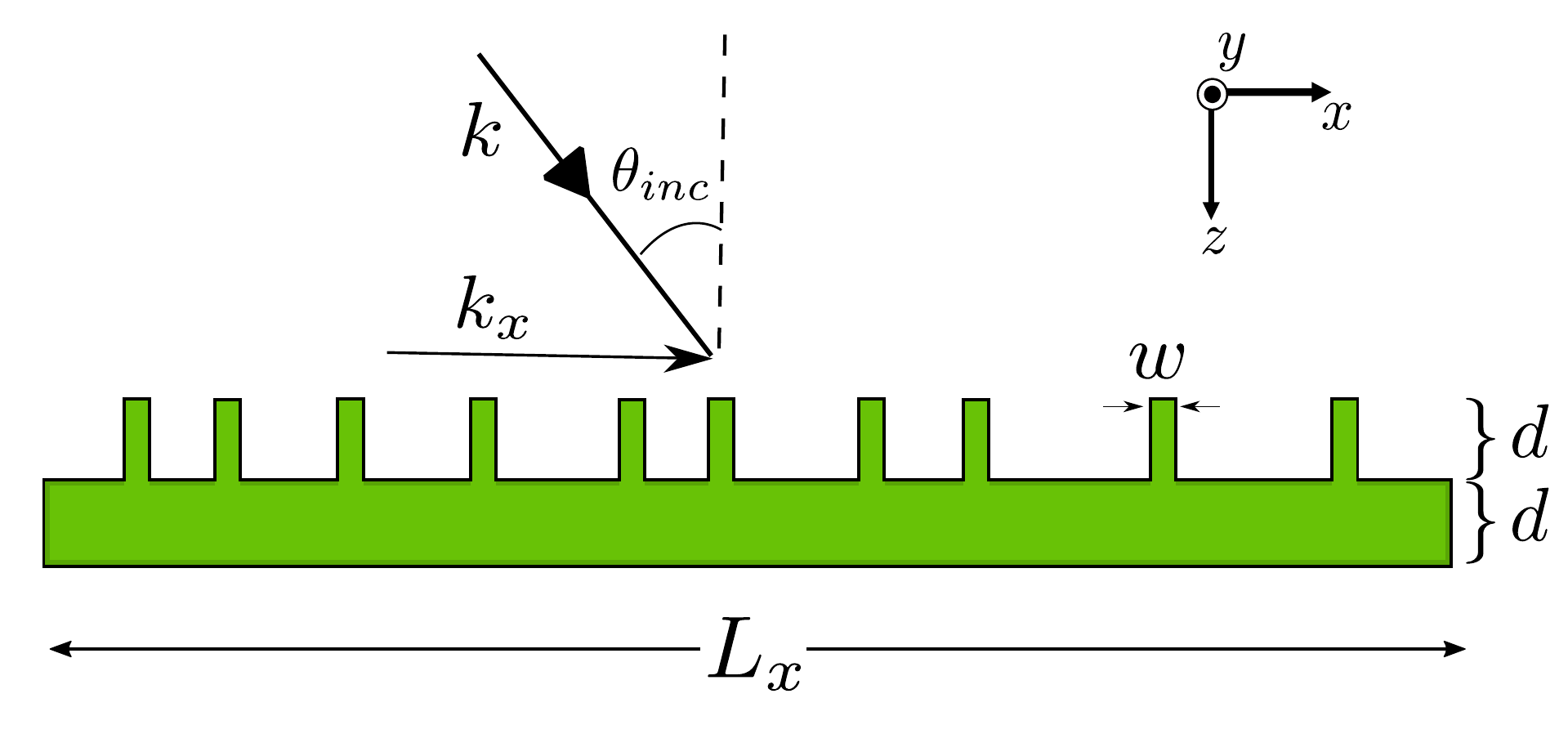}
	\caption{Schematic for the structure used in the sample FLAME-slab calculation. 
		The structure consists of 10 dielectric pillars positioned aperiodically on a dielectric slab.
		Light is incident from the top, with wavenumber $k$ and incidence angle $\theta_{\mathrm{inc}}$.}
	\label{fig:schematic}
\end{figure}

The general description of the problem in this section closely follows the recently published paper
\cite{Mansha-OpEx17}, which explores a new computational method, ``FLAME-slab,'' for  
electromagnetic wave scattering problems in aperiodic photonic structures -- specifically, 
structures possessing short-range regularity but lacking long-range order, 
such as amorphous or quasicrystalline lattices.  Structures of this type can exhibit a variety of interesting properties, e.g.
highly isotropic band gaps and fractal photonic spectra, but are difficult to 
study numerically \cite{Jin2001, Garcia2007, Florescu2009_2, Noh2011_1, Liang2013, 
	Mansha2016, Feng2005, Steurer2007, Liew2015, Knitter2015}.  
FLAME-slab exploits the short-range regularity of the structure by generating
a Trefftz basis in a relatively small segment of the structure.

As an example, we consider a slab substrate patterned with aperiodically placed 
but geometrically identical pillars (Fig.~\ref{fig:schematic}). 
The slab has thickness $d$, and there are 10 pillars 
of height $d$ and width $w = 0.8 \, d$.  Both the substrate and the pillars 
have dielectric constant $\varepsilon=12$.  The surrounding medium is air. 
In our calculations, we adopt computational units 
where the vacuum constants and the speed of light are all set to unity:
$\varepsilon_{0} = 1$, $\mu_{0} = 1$, $c = 1$. Then the frequency $f$ has the units 
of $1/\lambda$, where $\lambda$ is the free space wavelength.

Light is incident from the top, as shown in Fig.~\ref{fig:schematic}, 
with a wavenumber $k$ and incidence angle $\theta_{\mathrm{inc}}$ relative to the $z$-axis. 
We take the entire structure to be a supercell of length $L_{x}$, with quasi-periodic 
boundary conditions (see below).  The electric and magnetic fields in the structure 
are governed by Maxwell's equations:
\begin{align}
\begin{aligned}
\nabla \times \mathbf{E} &= ik\mathbf{H},\\
\nabla \times \mathbf{H} &= -ik\varepsilon\mathbf{E}.
\end{aligned}
\label{eq:Maxwell}
\end{align}
We consider the case where the electric field is $s$-polarized, $\mathbf{E}=E\hat{y}$, so that the magnetic field has the form $\mathbf{H}=H_{x}\hat{x}+H_{z}\hat{z}$. The quasi-periodic boundary conditions are:
\begin{align}
\begin{aligned}
\mathbf{E}(L_{x}/2,z) &= \mathbf{E}(-L_{x}/2,z)\exp(ik_{x}L_{x}), \\
\mathbf{H}(L_{x}/2,z) &= \mathbf{H}(-L_{x}/2,z)\exp(ik_{x}L_{x}),
\end{aligned}
\end{align}
where $k_x = k\sin \theta_{\mathrm{inc}}$ is the $x$-component of the incident wave vector $\bfk$.

The scattered electric field is defined as
\begin{equation}
   E_s(\mathbf{r}) = E_{\mathrm{tot}}(\mathbf{r}) - E_{\mathrm{inc}}(\mathbf{r}), \hspace{1cm} [\,\mathbf{r}\equiv (x,z)\,],
\end{equation}
where $E_{\mathrm{tot}}$ and $E_{\mathrm{inc}}$ are the total and incident electric fields, respectively. 
The magnetic field is split similarly. The scattered field is purely outgoing on both the upper side 
(towards the negative $z$-direction) and the lower side (towards the positive $z$-direction) of the structure.

\begin{figure}
	\centering
	\includegraphics[width=10cm]{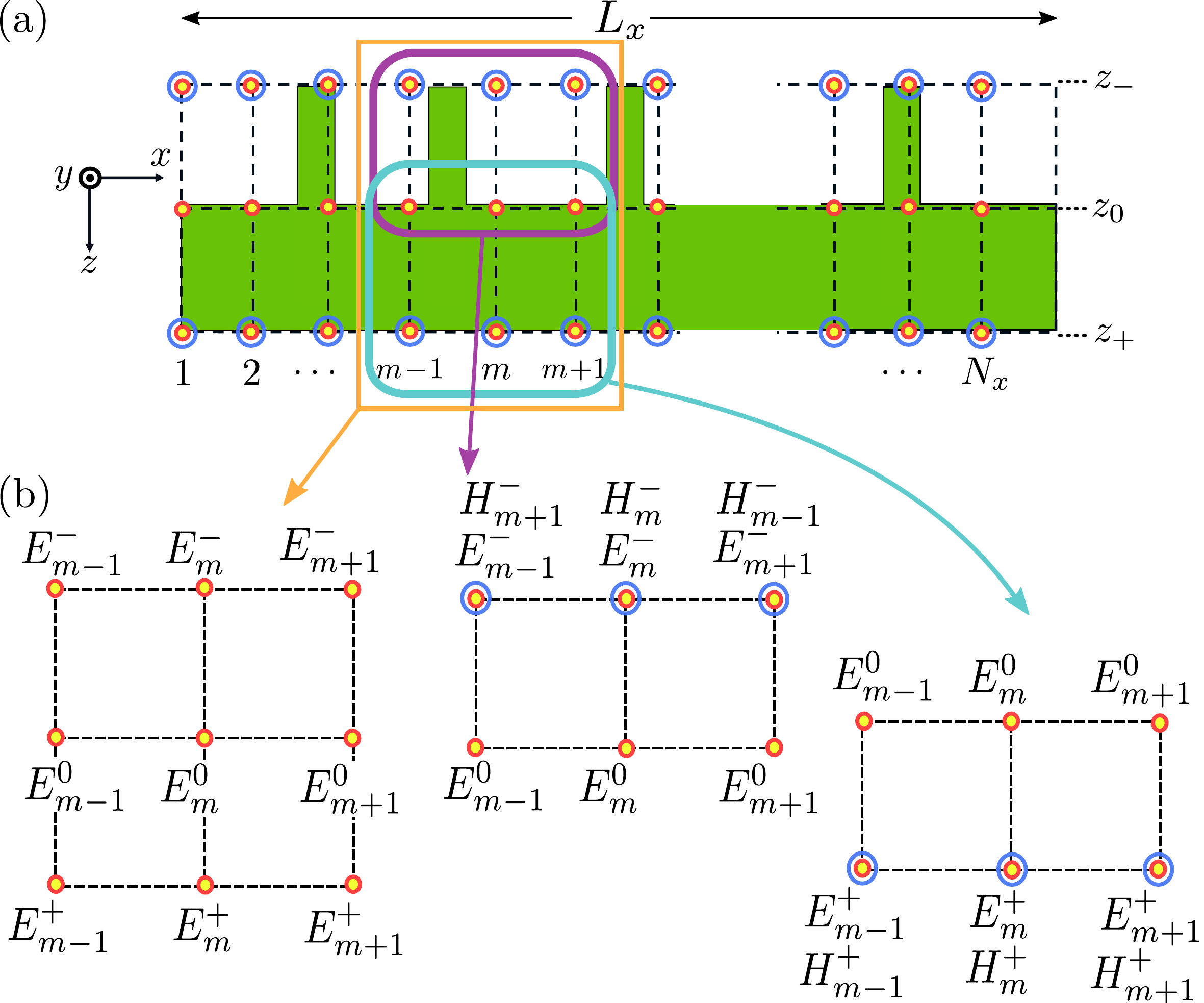}
	\caption{(a) Discretization of the structure into $N_x \times 3$ nodes ($N_x$ in the horizontal direction and 3 layers in the vertical direction). (b) Variations of the 9-point stencils.
		Left: 9 nodes with a single DoF (the values of the electric field). 
		Center and right: three nodes with a single DoF (the electric field) and another three nodes
		(double circles) with double DoF (electric and magnetic fields.)}
	\label{fig:discretization}
\end{figure}

Fig.~\ref{fig:discretization}(a) shows the discretization scheme for FLAME.  
The structure is discretized into $N_x$ grid points in the horizontal direction. In the vertical direction,
the number of layers is deliberately limited to three ($z_{-}$, $z_{0}$, $z_{+}$), 
to demonstrate that FLAME-slab can work well on very coarse grids. 
The electric fields in these three layers are denoted with $E^{\alpha}_m$, 
$\alpha=\{-,0,+\}$, $m = 1,2, \ldots, N_x$. 
Similarly, the magnetic fields in the upper and bottom layers 
are denoted with $H^{\beta}_m$, $\beta=\{-,+\}$.

We define three distinct types of patches, with their corresponding grid ``molecules''
and FD stencils. The first is a standard 9-point stencil containing just the electric field degrees of freedom (DoF), 
as shown in the left panel of Fig.~\ref{fig:discretization}(b).  The second is a 6-point stencil over the middle and top layers, 
containing both the electric and magnetic fields (middle panel).  The third is a 6-point stencil over the middle and bottom layers, 
containing both the electric and magnetic fields (right panel of Fig.~\ref{fig:discretization}(b)).

Each type of patch thus contains 9 degrees of freedom. FLAME uses 8 basis functions, 
to be determined by solving Maxwell's equations for ``Trefftz cells'' matching 
the local dielectric environment in each patch.  Each Trefftz cell contains 
a segment of length $L_{i}$ with a single pillar on the substrate; 
quasi-periodic boundary conditions are imposed.  We choose $L_{i} \ll L_x$, 
so that Maxwell's equations can be solved much more rapidly for the Trefftz cell 
than for the entire aperiodic structure.  We generate 8 different Trefftz basis functions 
by picking two different segment lengths ($L_1$ and $L_2$), and four different angles 
of incidence for each $L_i$.  To compute the fields in the Trefftz cell, we use the existing rigorous coupled wave analysis (RCWA) solver $S^{4}$ \cite{Liu2012}.

The FLAME procedure now yields a matrix equation of the form
\begin{equation}
    \mathbf{A}_{\mathrm{FL}} \, \boldsymbol{\psi}_{tot} = 0,
\label{eq:Aflame_psi}
\end{equation}
where $\mathbf{A}_{\mathrm{FL}}$ is a matrix of stencil coefficients and $\boldsymbol{\psi}_\mathrm{tot}$ 
is a column vector containing the nodal values of the total electric and magnetic fields. In our examples,
$\mathbf{A}_{\mathrm{FL}}$ has the size $3N_{x}\times 5N_{x}$, and $\boldsymbol{\psi}_\mathrm{tot}$ 
has the size $5N_{x}\times 1$; we emphasize that this is just one possible choice of discretization, and other choices can be handled 
in a completely analogous way.  Details about the calculation of $\mathbf{A}_{\mathrm{FL}}$ 
can be found  in \cite{Mansha-OpEx17}.  

FLAME schemes need to be supplemented with radiation boundary conditions. 
One way of implementing such conditions is via the Dirichlet-to-Neumann (DtN) maps
in the semi-infinite air strips above and below the slab. DtN maps
can be efficiently calculated via Fast Fourier Transforms (FFTs). 
More specifically, from Maxwell's equations in free space,
\begin{equation}
   H_s(x,z) = \frac{i}{\omega} \frac{\partial E_s}{\partial z}
\label{eq:EandHMaxwell}
\end{equation}
The operating frequency $\omega=2\pi f = k$, under the assumed normalization $c=1$.  
We expand the scattered electric field into its Fourier series:
\begin{equation}\label{eq:EscatBloch}
   E_s(x,z) = \sum_{n} c_{n} \exp[i(k_{nz}z + k_{nx}x)] \exp(iqx),
\end{equation}
where the factor of $\exp(iqx)$ comes from the quasiperiodic boundary conditions in the $x$ direction, 
with $q = k\sin \theta_{\mathrm{inc}}$. The summation $n$ runs over the integer values, $k_{nx}=2\pi n/ L_x$ 
is the horizontal wavenumber, and
\begin{equation}
k_{nz} = \pm \sqrt{k^2 - (k_{nx} + q)^2}.
\label{eq:knzexpression}
\end{equation}
In the above equation, the choice of $\pm$ depends upon the layer we are dealing with 
($-$ for the upper layer and $+$ for the bottom layer), so that the scattered field is outgoing.  
Eqs.~\eqref{eq:EandHMaxwell} and \eqref{eq:EscatBloch} give
\begin{equation}\label{eq:HscatBloch}
   H_s(x,z) = -\frac{1}{\omega}\sum_{n} c_{n} \, k_{nz} \, \exp\left[i(k_{nx}x + k_{nz}z)\right] \exp(iqx).
\end{equation} 
%

The coefficients $c_n$ in \eqref {eq:EscatBloch} can be efficiently computed
via a Fast Fourier Transform, and then the scattered magnetic field \eqref{eq:HscatBloch}
can be obtained via the respective inverse transform (detailed expressions can be
found in \cite{Mansha-PhDthesis18}). This leads to equations in the following matrix form:
\begin{equation}\label{eq:Adef}
    \begin{pmatrix} 
       \mathbf{A}_{\mathrm{FL}} \\ \mathbf{A}_{\mathrm{BC}} 
    \end{pmatrix} 
    \boldsymbol{\psi}_s
    = 
    \begin{pmatrix} 
       -\mathbf{A}_{\mathrm{FL}} \boldsymbol{\psi}_{\mathrm{inc}} \\ \boldsymbol{0}
     \end{pmatrix},
\end{equation}
where $\mathbf{A}_{\mathrm{FL}}$ is a sparse sub-matrix obtained using FLAME, and $\mathbf{A}_\mathrm{BC}$ 
is sub-matrix obtained from the boundary relations \cite{Mansha-OpEx17}.  
In our 2D examples, standard direct solvers in Matlab were sufficient for finding $\boldsymbol{\psi}_s$. 
In 3D, iterative solvers will need to be used,
but this issue is completely beyond the scope of the present paper.

\subsection{Results}

\begin{figure}
	\centering
	\includegraphics[width=0.9\textwidth]{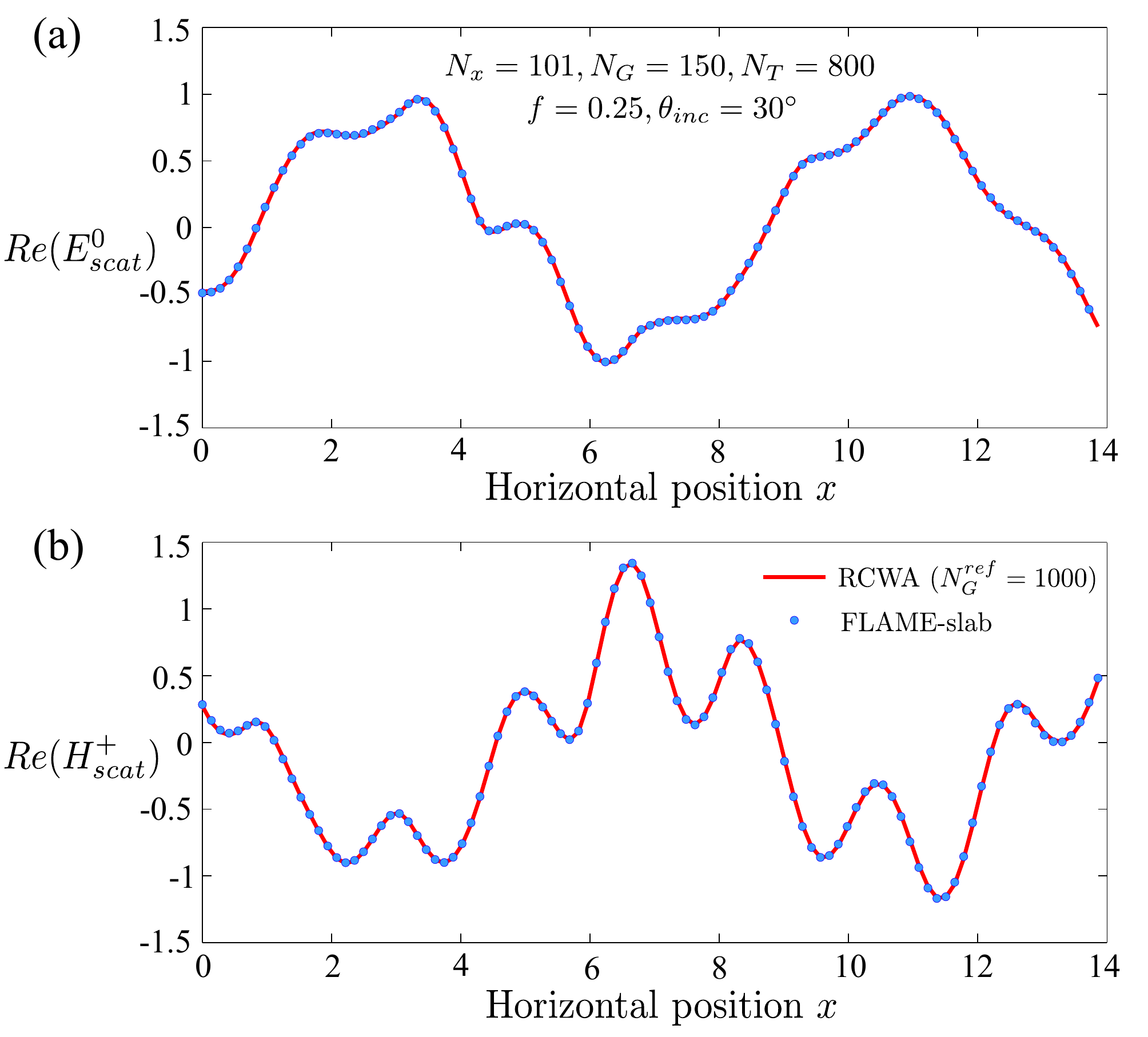}
	\caption{(a) Real part of the scattered electric field $E_{scat}$ in the middle layer ($z_0$). (b) Real part of the scattered magnetic field $H_{scat}$ in the bottom layer ($z_{+}$). The calculations were done for the slab shown in Fig.~\ref{fig:schematic}, with $f=0.25$ and $\theta_{\mathrm{inc}}=30\degree$. The FLAME-slab parameters are $N_x=101, $$N_G=150$, and $N_T=800$. Blue dots show the FLAME-slab results and the red curve shows the result from RCWA obtained by setting $N_{G}^{ref}=1000$.}
	\label{fig:EHfields}
\end{figure}

Fig.~\ref{fig:EHfields} compares the fields calculated using FLAME-slab to a reference RCWA calculation.  
The structure is the one shown in Fig.~\ref{fig:schematic}, with frequency $f=0.25$ and incidence angle 
$\theta_{\mathrm{inc}} = 30\degree$.  For the FLAME-slab calculation, we take a horizontal discretization of $N_x=101$, 
and precompute the Trefftz basis functions with $N_G=150$ (the number of expansion terms used in the RCWA subroutine \cite{Liu2012}) 
and $N_T=800$ (the cell dicretization used for storing the Trefftz basis functions).  The pure RCWA reference solution 
is computed using $N_{G}^{ref}=1000$ -- an ``overkill'' setting meant to produce a highly accurate solution. 
 The figure shows two representative field components: the real part of the scattered electric field ($E^{0}_s$) 
 in the middle layer ($z_0$) in Fig.~\ref{fig:EHfields}(a), and the scattered magnetic field ($H^{+}_s$) 
 in the bottom layer ($z_{+}$) in Fig.~\ref{fig:EHfields}(b). The FLAME-slab solution is seen to be
in excellent agreement with the RCWA solution.

\begin{figure}
	\centering
	\includegraphics[width=0.9\textwidth]{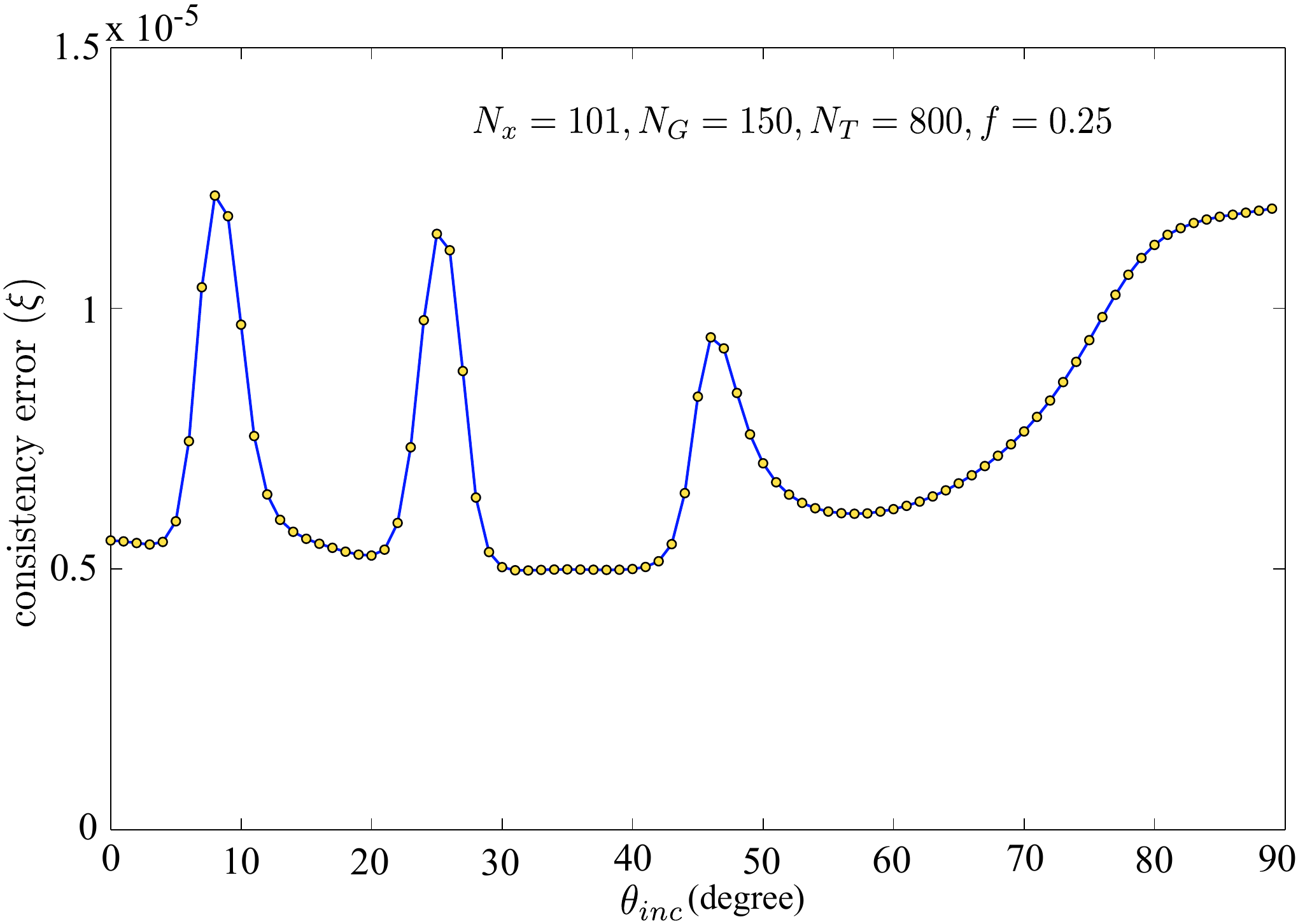}
	\caption{Consistency error ($\xi$) vs. angle of incidence $\theta_{\mathrm{inc}}$ 
		for the 10 pillar system as shown Fig.~\ref{fig:schematic}. The value of the parameters used are: 
		$N_x=101$, $N_G=150$, $N_T=800$ and $f=0.25$.}
	\label{fig:cons_error_vs_theta}
\end{figure}


The central issue of this paper is approximation, and the finite-difference measure 
most closely related to it is the (normalized) consistency error
\begin{equation}
   \xi = \frac{\| \mathbf{A}_{\mathrm{FL}} \, 
   \boldsymbol{\psi}_{\mathrm{tot}}^{\mathrm{ref}} \|}
   {\| \mathbf{A}_{\mathrm{FL}} \| \, 
   	\|\boldsymbol{\psi}_{\mathrm{tot}}^{\mathrm{ref}} \|}.
\label{eq:consistency}
\end{equation}
where Euclidean vector norms and the Frobenius matrix norm are implied.

In \eqref{eq:consistency}, $\boldsymbol{\psi}_{\mathrm{tot}}^{\mathrm{ref}}$
should ideally be the exact solution, which is not available; hence
an overkill RCWA solution with $N_{G}^{ref}=1000$ is used in its stead. 

Since FLAME-slab contains a few adjustable parameters, 
we study the dependence of the consistency error on these parameters separately.

Fig.~\ref{fig:cons_error_vs_theta} displays the consistency error versus the incidence angle $\theta_{\mathrm{inc}}$.  For this calculation, we set $N_x=101$, $N_G=150$, $N_T=800$ and $f=0.25$.  The consistency error oscillates but remains bounded by $\lesssim 10^{-5}$ over the entire range of $\theta_{\mathrm{inc}}$.

\begin{figure}
	\centering
	\includegraphics[width=10cm]{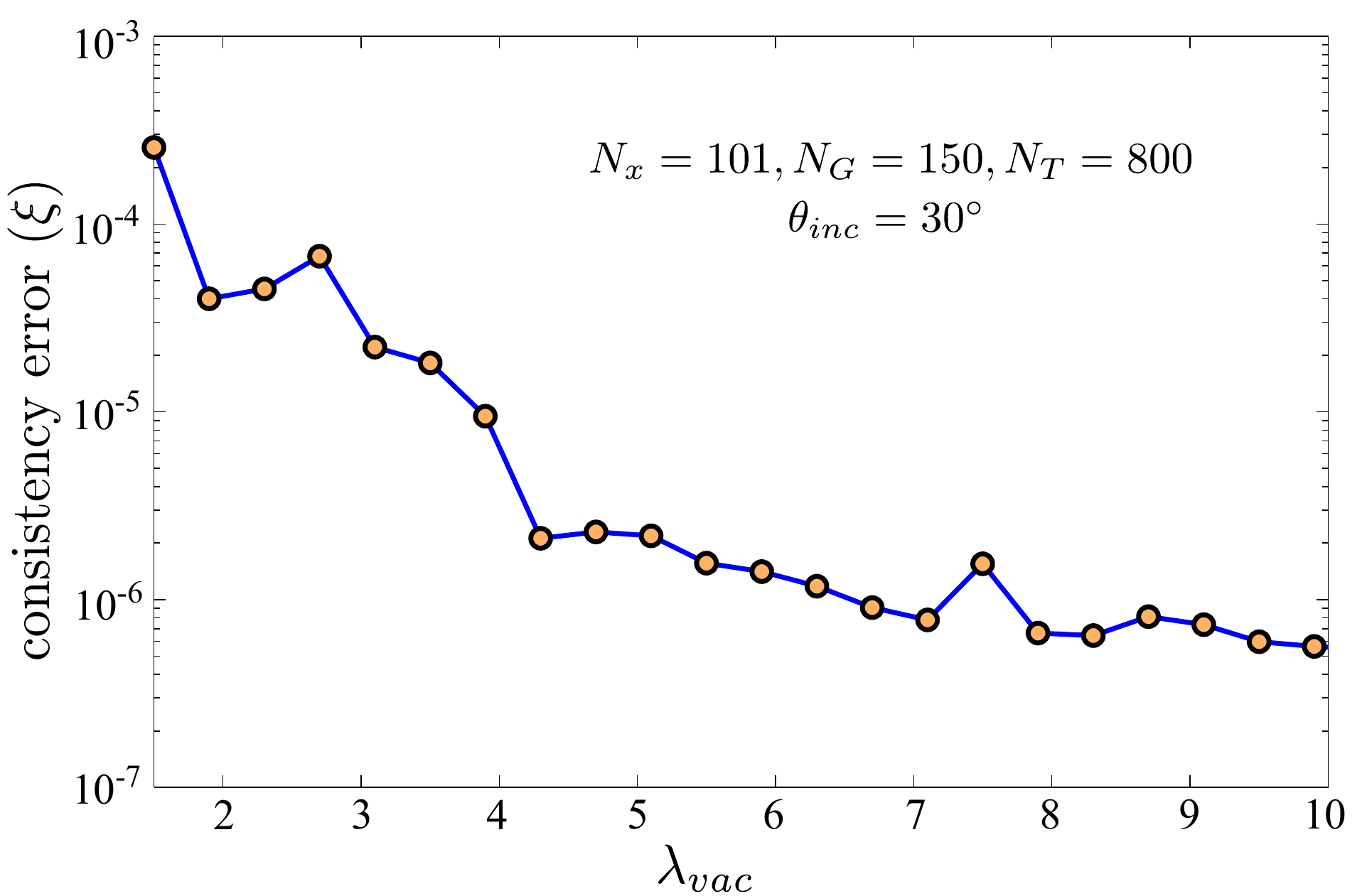}
	\caption{Consistency error ($\xi$) vs. $\lambda_{\mathrm{vac}}$ (vacuum wavelength) for the slab structure shown in Fig.~\ref{fig:schematic}, with $\theta_{\mathrm{inc}} = 30\degree$. The FLAME-slab parameters are 
		$N_x=101$, $N_G=150$, and $N_T=800$.}
	\label{fig:cons_error_vs_lambda}
\end{figure}

\begin{figure}
	\centering
	\includegraphics[width=10cm]{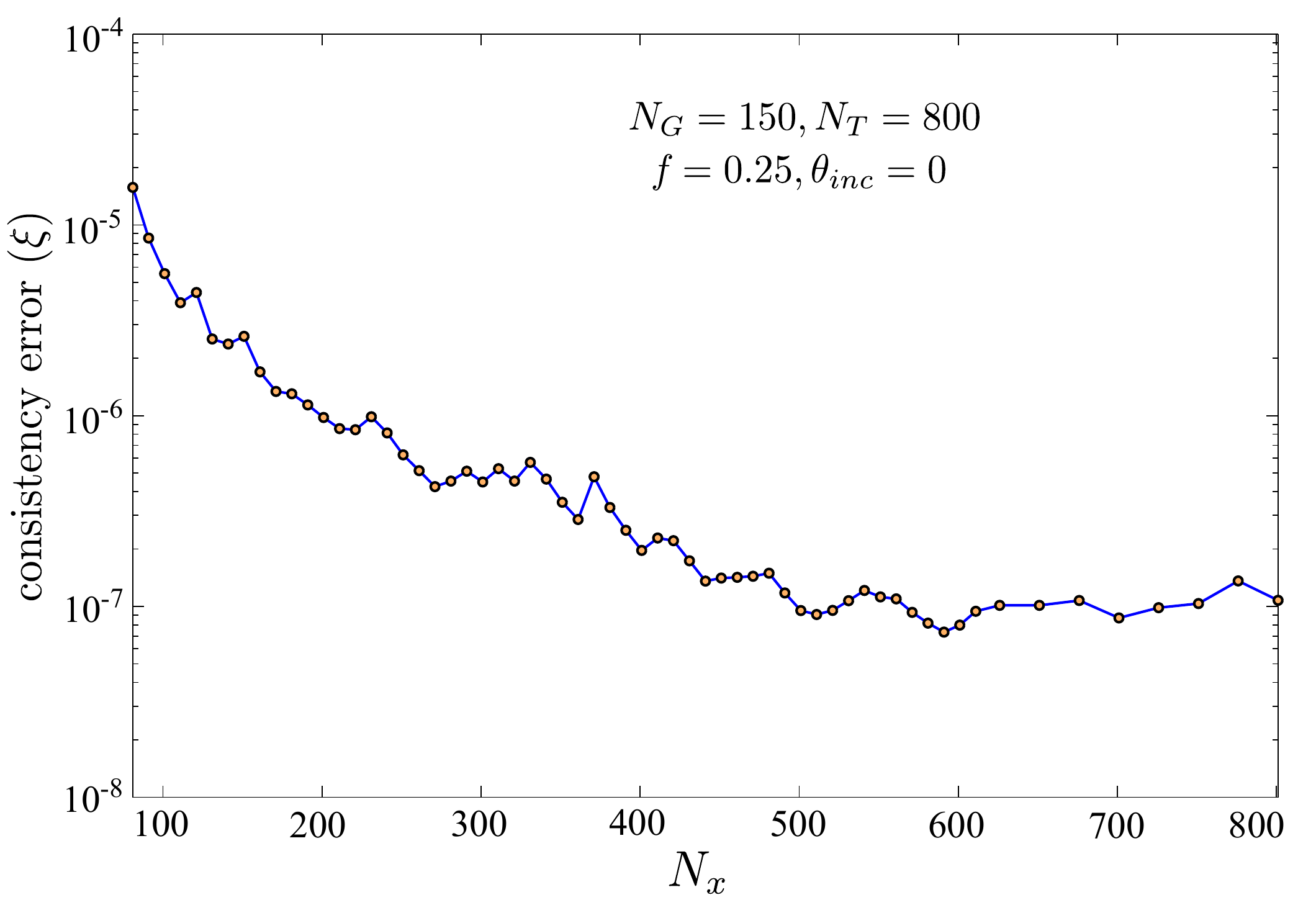}
	\caption{Consistency error ($\xi$) vs. $N_x$ for the slab shown in Fig.~\ref{fig:schematic}, with $f = 0.25$ 
		and $\theta_{\mathrm{inc}} = 0$. The FLAME-slab parameters are $N_G=150$ and $N_T=800$.}
	\label{fig:cons_error_vs_Nx}
\end{figure}

Fig.~\ref{fig:cons_error_vs_lambda} shows the consistency error versus the vacuum wavelength 
$\lambda_{\mathrm{vac}}$ for the 10 pillar system, with fixed incidence angle $\theta_{\mathrm{inc}} = 30\degree$. 
The FLAME-slab parameters are fixed at $N_x=101$, $N_G = 150$, and $N_T = 800$.  
As $\lambda_{\mathrm{vac}}$ is increased, $\xi$ decreases from $10^{-4}$ to around $10^{-6}$.  
Past this point, $\xi$ saturates.

Fig.~\ref{fig:cons_error_vs_Nx} shows the consistency error versus spatial discretization $N_x$, 
for $f = 0.25$ and normal incidence $\theta_{\mathrm{inc}}=0$.  The other FLAME-slab parameters 
are $N_G=150$ and $N_T=800$.  The consistency error decreases with $N_x$, saturating at $\approx 10^{-7}$ for $N_x \gtrsim 500$.

For the purposes of the paper, the main qualitative conclusion of this section
is that \textit{Trefftz functions}, on which FLAME-slab is based, 
\textit{provide an accurate approximation
of the electromagnetic field in a geometrically and physically complex structure}.

\section{The Accuracy of Trefftz Approximations}
\label{sec:Trefftz-approximation}
%
\subsection{An Interpolation Argument}\label{sec:Interpolation-argument}
The numerical results for the two application examples of
the previous sections show that Trefftz approximations
are surprisingly effective. What explains their high accuracy?

As noted in \sectref{sec:Intro}, in the mathematical literature
this question has been studied primarily for homogeneous subdomains 
(harmonic polynomials, plane/cylindrical/spherical wave expansions) 
but needs to be posed much more broadly,
because complex inhomogeneous media are of great theoretical
and practical interest. This section is an attempt to understand
the general mechanisms of high accuracy of Trefftz approximations.
Due to the complexity of this subject, some of the material,
especially that of \sectref{sec:Random-matrices}, is speculative and intended to
stimulate further analysis and discussion.

In the case of Trefftz homogenization (Section~\ref{sec:Trefftz-homogenization}),
one can apply an interpolation argument using the summary in Section~\ref{sec:Trig-interpolation}.
Indeed, the key parameters in our homogenization methodology are the
boundary averages of the Bloch fields \eqref{eqn:EH0-eq-face-avrg-eh}.
Each of these averages is, trivially, a periodic function of the angle
(direction) of propagation of the respective Bloch wave and, as such,
can be accurately approximated by the trigonometric interpolant
over a set of equispaced knots. But these knots correspond precisely to
the basis set of Bloch waves chosen in our procedure.
Per \sectref{sec:Trefftz-homogenization}, the accuracy of this interpolation is
$\mathcal{O}(N^{l+1})$ if the respective Bloch average is $l$ times continuously differentiable,
or, under additional analyticity assumptions, even $\mathcal{O}(\exp(-\alpha N))$,
where $N$ is the size of the Bloch basis set (which is the same as the number of interpolation knots).

\begin{figure}
	\centering
	\includegraphics[width=0.65\linewidth]{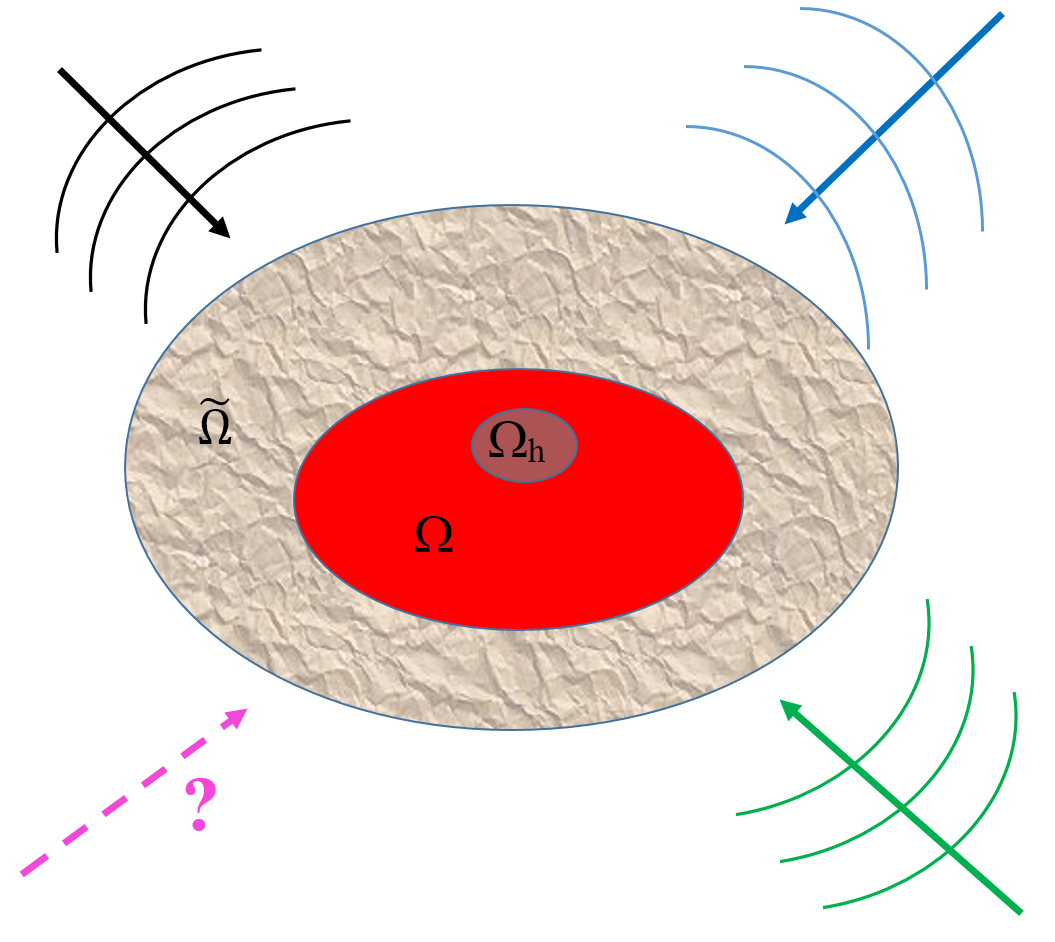}
	\caption{An inhomogeneous scatterer $\Omegarm$ (solid red) is enclosed 
		in a shell $\tilde{\Omegarm}$ (textured). The material parameters
		are fixed within $\Omega$ in all cases. However, in $\tilde{\Omegarm}$
		these parameters for the unknown field (dashed arrow) may differ from
		the parameters used in the construction of the Trefftz training set
		(solid arrows). A local Trefftz approximation of the unknown field
		is sought in a small subdomain $\Omega_h \subset \Omega$.}
	\label{fig:test-waves-random}
\end{figure}

In our second example of wave propagation and scattering in a slab geometry,
the interpolation argument is not sufficient. This is because our Trefftz functions
are defined over \textit{a segment} of the structure, 
whereas the full electromagnetic problem
is defined over the \textit{whole structure}. Hence a more sophisticated explanation
for the accuracy of Trefftz approximations in this case is needed.

We start with a slightly more abstract physical setup than that of Fig.~\ref{fig:test-waves}.
Namely, let us assume, as before, that an inhomogeneous scatterer occupies a Lipschitz domain 
$\Omegarm$ (solid red in Fig.~\ref{fig:test-waves-random}) which is enclosed 
in a shell $\tilde{\Omegarm}$ (textured area). As previously,
we consider a Trefftz ``training set'' corresponding to several incident waves, 
and are interested in approximating a different, generally unknown, solution 
in a small subdomain $\Omegarm_h \subset \Omegarm$.
This approximation can be used, for example, to generate a high-order difference scheme
in $\Omegarm_h$, as was done in \sectref{sec:FLAME-slab}. 

The Trefftz training set is also generated
for fixed position-dependent parameters in $\Omegarm \cup \tilde{\Omegarm}$. 
Importantly, however, the unknown solution may correspond to material parameters 
which \textit{differ} in $\tilde{\Omegarm}$ 
from those assumed for the training set (but are the same in $\Omegarm \supset \Omegarm_h$).
The presence of the variable layer $\tilde{\Omega}$ makes this case peculiar. 
The following section examines why accurate \textit{local} Trefftz approximations can still be expected.

\subsection{An Auxiliary ``Reference'' Basis}\label{sec:Auxiliary-basis}
%
Let us assume that in $\Omega_h$ there is \textit{an auxiliary basis}
$\zeta_\alpha$ ($\alpha = 1,2, ..., n_\zeta$) which can provide an accurate
approximation of a (generic) solution of the wave equation:
\begin{equation}\label{eqn:E-vs-modes}
     u(\mathbf{r}) = \sum_{\alpha}
     \gamma_\alpha \zeta_\alpha (\mathbf{r}) + \delta(\bfr),
     ~~ \bfr \in \Omega_h
\end{equation}    
$$
    \| \underline{\gamma} \|_2 \equiv \| \{\gamma_\alpha \} \|_2
     \leq C(\Omega_h, n_\zeta, k) \| u \|_{H^1(\Omega_h)},
     ~~ \| \delta \|_{H^1(\Omega_h)} \leq c(\Omega_h, n_\zeta, k) \| u \|_{H^1(\Omega_h)}
$$
Here $\delta$ is an error term, $\underline{\gamma}$ is a coefficient
vector, $C$ and $c$ are some generic constants, the latter being ``small''
in some sense (see Theorems below). In the specific example of $s$-wave
scattering in Section~\ref{sec:FLAME-slab}, the unknown is the $E$-field; 
but here we use the ``generic'' symbol $u$ 
as an indication that our analysis could be applied more broadly.

Assuming that \eqref{eqn:E-vs-modes} holds, one applies it to the training 
set $\underline{\psi}_T$ of $n_T$ Trefftz waves, and arrives at the linear transformation
$$
   \underline{\psi}_T(\bfr) = P_{\zeta \rightarrow T} \underline{\zeta}(\bfr) 
   + \underline{\delta}_T(\bfr)
$$
where column vectors are underlined;  $P_{\zeta \rightarrow T}$
is the $ n_T \times n_\zeta $ transformation matrix,
and $\underline{\delta}_T$ is the approximation error for the Trefftz functions
in terms of the local $\zeta$ basis.
If $n_\zeta \leq n_T$, and if matrix $P_{\zeta \rightarrow T}^* P_{\zeta \rightarrow T}$
is invertible, then
$$
   \underline{\zeta}(\bfr) = P_{\zeta \rightarrow T}^+ \underline{\psi}_T(\bfr) + 
   P_{\zeta \rightarrow T}^+ \underline{\delta}_T(\bfr)
$$
where the `+' subscript indicates the Moore-Penrose pseudoinverse.

\begin{figure}\label{fig:Trefftz-vs-reference-basis}
	\centering
	\includegraphics[width=0.4\linewidth]{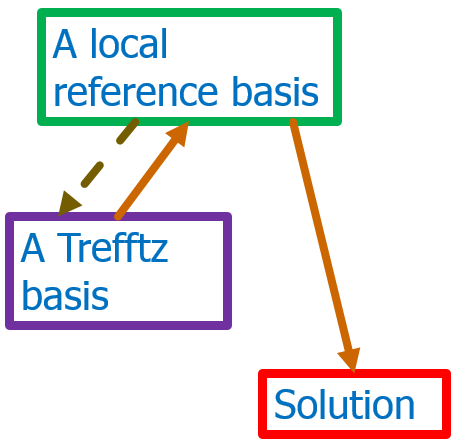}
	\caption{A schematic illustration of the role of the reference basis.
		If the Trefftz basis and the solution of a given boundary value problem
		can be approximated via the reference basis, and if the reference-to-Trefftz
		transformation has a bounded pseudoinverse, then one can approximate
		the solution via the Trefftz basis (by following, conceptually,
		the two solid arrows in the sketch).
	}
\end{figure}

The solution in $\Omegarm_h$ can therefore be expressed as
\begin{equation}\label{eqn:Eexact-Trefftz-approximation}
    u(\bfr) = \underline{\gamma}^T \underline{\zeta}(\bfr) = 
    \underline{\gamma}^T P_{\zeta \rightarrow T}^+ \underline{\psi}_T(\bfr) + 
    \underline{\gamma}^T P_{\zeta \rightarrow T}^+ \underline{\delta}_T(\bfr)
   +  \delta_u(\bfr)
\end{equation}
where $\delta_u$ is the approximation error of this solution via the $\zeta$ basis.
Thus the smallness of the Trefftz approximation error hinges on the smallness of the norm of 
the pseudoinverse $P_{\zeta \rightarrow T}^+$ -- that is, on the inverse
of its minimum singular value $\sigma_{\min}$; we discuss that below.

The transformations above are schematically illustrated in Fig.~\ref{fig:Trefftz-vs-reference-basis}.
If the Trefftz basis and the solution $u$
can be approximated via the reference basis as in \eqref{eqn:E-vs-modes}, 
and if $\sigma_{\min} (P)$
is bounded from below, then one can approximate
$u_{\mathrm{exact}}$ via the Trefftz basis (by following, conceptually,
the two solid arrows in the sketch).

An example of this auxiliary basis is, in the special case of a homogeneous
domain $\Omega_h$, a set of cylindrical harmonics $\zeta_{\mathrm{cyl}}(r,k,\theta,n)
= J_n (kr) \exp(in \theta)$,
$n = 0, \pm 1, \pm 2, \ldots$; $J_n$ is the Bessel function of the first kind.
Detailed error analyses have been carried out by Melenk, Hiptmair, Moiola
and Perugia \cite{Melenk95,Hiptmair2016,Moiola-PhD11}. For our purposes,
the most convenient final results can be found in \cite{Babuska97,Melenk99}.

\textbf{\cite[Theorem 4]{Babuska97}.} Let $\Omegarm \subset \mathbb{R}^2$ be a simply connected, 
bounded Lipschitz domain. Let $\tilde{\Omegarm} \supset \Omegarm$ and
assume that $u \in L^2(\tilde{\Omegarm})$ solves the homogeneous Helmholtz equation on $\tilde{\Omegarm}$. ¹
Then
\begin{equation}\label{eqn:Babuska-Melenk-cyl-harm-estimate1}
    \inf_{u_p \in V_p} \| u - u_p \|_{H^1(\Omegarm)}  \,\leq\, C \exp(-\gamma p) 
    \, \| u \|_{L^2{(\tilde{\Omegarm})}}
\end{equation}
where $V_p \equiv \mathrm{span} \{\zeta_{\mathrm{cyl}}(r,k,\theta,n) \}$,
$n = 0,1, \ldots, p$; $C, \gamma$ depend only on $\Omegarm$, $\tilde{\Omegarm}$, and 
the wavenumber $k$.

Under the assumptions of this theorem, the presence of a ``buffer region'' 
$\tilde{\Omegarm} - \Omegarm$ ensures that high-order harmonics from the boundary
of $\tilde{\Omegarm}$ die out sufficiently. If this assumption is not made,
an alternative error estimate, dependent on the level of smoothness of
the solution, reads:

\textbf{\cite[Theorem 5]{Babuska97}.} Let $\Omega \subset \mathbb{R}^2$ be a simply connected, 
bounded Lipschitz domain, star-shaped with respect
to a ball. Let the exterior angle of $\Omegarm$ be bounded from below by $\lambda \pi$,
$0 < \lambda < 2$. Assume that $u \in H^s(\Omegarm)$, $s > 1$, satisfies the homogeneous 
Helmholtz equation. Then\footnote{There is an apparent misprint in \cite{Babuska97}:
	$H^k$ instead of $H^s$ in the norm on the right hand side.}
\begin{equation}\label{eqn:Babuska-Melenk-cyl-harm-estimate2}
   \inf_{u_p \in V_p} \| u - u_p \|_{H^j(\Omegarm)}  \,\leq\, 
    C_j \, \left( \frac{\ln^2 p}{p} \right)^{\lambda(s-j)}
    \, \| u \|_{H^s{(\tilde{\Omegarm})}},
    ~~~ j = 0,1, \ldots, [s]
\end{equation}
Obviously, in our case $\Omegarm_h$ plays the role of the generic $\Omegarm$ in the
estimates above. These estimates of the error term
$\delta$ in \eqref{eqn:E-vs-modes} are valid for the 2D Helmholtz equation in
a physically \textit{homogeneous medium} within $\Omegarm_h$.

Also in the special case of a homogeneous domain $\Omega_h$,
and the Trefftz set consisting of plane waves traveling in $n_T$ equispaced
angular directions, the norm of the pseudoinverse $P_{\mathrm{cyl} \rightarrow \mathrm{PW}}^+$
can be evaluated explicitly.
From the Jacobi-Anger expansion, the entries of
the matrix $\hat{P} \equiv P_{\mathrm{cyl} \rightarrow \mathrm{PW}}$ are
\begin{equation}\label{eqn:P-cyl-to-PW}
  \hat{P}_{ml} \,=\, i^l \exp \left(-iml \, \frac{2\pi}{n_T} \right),
  ~~ 0 \leq m \leq n_T-1, ~~ 0 \leq l \leq n_\zeta - 1
\end{equation}
This matrix corresponds to a discrete Fourier transform, and its
columns are easily shown to be orthogonal, so that
\begin{equation}\label{eqn:P-P-diag}
    \hat{P}^* \hat{P} \,=\, n_T I_{n_\zeta},
    ~~~ n_\zeta \leq n_T
\end{equation}
where $I_{n_\zeta}$ is the identity matrix of dimension $n_\zeta$.
It then immediately follows that
\begin{equation}\label{eqn:sigma-min-P-cyl-PW}
   \| P_{\mathrm{cyl} \rightarrow \mathrm{PW}}^+ \|_2 \,=\,
   \sigma_{\min}^{-1}( P_{\mathrm{cyl} \rightarrow \mathrm{PW}}) = n_T^{-\frac12}
\end{equation}
so in this case stability of the transformation is guaranteed.

\subsection{A Connection with Random Matrix Theory}\label{sec:Random-matrices}
%
A natural, and critical, question is whether the well-posedness of the transformation noted above 
is accidental and valid in special cases only, or whether it has broader applicability.
Practical experience with multiparticle problems, random and quasi-random structures 
of different kind \cite{Tsukerman05,Tsukerman06,Tsukerman-PBG08,Dai-Webb11,Mansha-OpEx17}, 
\cite[Chapters 4, 6]{Tsukerman-book07} strongly suggests the latter.
Rigorous mathematical analysis is so far available only for a narrow subset of cases
\cite[Chapter IV]{Melenk95}, \cite[Section 3]{Melenk99},
\cite{Laghrouche-jumps2005,Imbert-Gerard-interp-PW2015}, and may constitute an interesting
direction of future research.

In the remainder of this section, we outline -- on physical grounds --
a curious connection between the accuracy of Trefftz approximations 
and the theory of random matrices. This theory dates back
to von Neumann and Wigner \cite{vonNeumann-inverting-matrices-1947,Wigner-random-matrices-1955} and is now quite mature
\cite{Akemann-handbook-random-matrices-2011,Tao2010,
	Rudelson-Vershynin-least-sing-value2008,Rudelson-Vershynin-smallest-sing-value2009,
	Rudelson-Vershynin-nonasymptotic-sing-value2010}.
Particularly relevant to us is the following result.

\vskip 0.1in

\textbf{Rudelson \& Vershynin \cite[Theorem 3.3]{Rudelson-Vershynin-nonasymptotic-sing-value2010}}.\\
	Let $A$ be an $N \times n$ random matrix whose entries are independent
	and identically distributed (i.i.d.) subgaussian random variables
	with zero mean and unit variance. Then
	$$
	P \left( \sigma_{\min}(A) \le \epsilon (\sqrt{N} - \sqrt{n-1}) \right) 
	\le (C\epsilon)^{N-n+1} + c^N,
	~~~ \epsilon \ge 0
	$$
	where $C > 0$  and $c \in (0,1)$ depend only on the subgaussian
	moment of the entries.

\vskip 0.1in

The connection of this theorem with the previous
subsection can be outlined as follows.
\begin{itemize}
	\item The Trefftz ``training set'' can be viewed as a particular realization
	of some random distribution (e.g. angles of incidence randomly chosen
	and/or random properties of the ``shell'' $\tilde{\Omega}$).
	A notable feature of random matrix theory is \textit{universality}:
	only mild dependence of the spectral bounds on the distribution of
	the random variables. 
	\item
	One major restrictive condition, however, is that the matrix entries be i.i.d. variables.
	Strictly speaking, this condition can be immediately ascertained only under additional symmetry assumptions,
	e.g. the bases being invariant under rotation by a given angle. It is hoped that such strong assumptions
	can be relaxed. 
	\item
	The assumption of zero mean is less restrictive and valid if the probability
	distribution of each function in the Trefftz training set $\underline{\psi}_T(\bfr)$,
	for all $\bfr$, is invariant with respect to
	the sign change of that function.
	\item Clearly, the theorem is applied with $N \equiv n_T$, $n \equiv n_\zeta$.
	\item The assumption that the distribution is subgaussian is satisfied,
	in particular, by all bounded random variables and hence is not restrictive.\footnote{A random variable $X$ is called subgaussian if
		there exists a positive constant $w$ such that
		$P (|X| > x) \le 2 \exp(-x^2/w^2)$ for $x > 0$.}\footnote{
		There is an apparent misprint in \cite{Rudelson-Vershynin-nonasymptotic-sing-value2010}:
		$n \times n$ instead of $N \times n$.}
	\item Complex bases and matrices can be decomplexified by the
	substitutions of the form $\psi \rightarrow (\mathrm{Re} \, \psi, \mathrm{Im} \, \psi)^T$,
	$P \rightarrow 
	\begin{pmatrix}
	\mathrm{Re} \, P & -\mathrm{Im} \, P \\
	\mathrm{Im} \, P & \mathrm{Re} \, P 
	\end{pmatrix}
	$.
	This preserves the relevant norms and hence does not affect the spectral bounds.
	\item The assumption of unit variance is obviously a matter of scaling only.
\end{itemize}

The theorem affirms that stability \eqref{eqn:sigma-min-P-cyl-PW} 
of the transformation is not accidental. In fact,  with a probability close to one,
$\sigma_{\min}(P)$ is not small, for any reasonable choice of the Trefftz basis.

\section{Conclusion}
\label{sec:Conclusion}
%
The key argument of this paper is that Trefftz approximations --
that is, approximations by functions satisfying (locally) a given
differential equation -- deserve to be studied and applied
more broadly than is traditionally done. Conventionally, these approximations
are used in homogeneous subdomains, where the underlying differential equation
has constant coefficients; this is done in various contexts (GFEM, DG, FD).

As an illustration of a much broader use of Trefftz functions,
the paper reviews two disparate but representative examples:
(i) non-asymptotic and nonlocal two-scale homogenization of periodic
electromagnetic media, and (ii) special Trefftz FD (FLAME) schemes
for wave scattering from photonic structures with slab geometries.
In both cases, Trefftz approximations are applied in complex inhomogeneous
domains and prove to be quite effective.

We discuss possible mechanisms engendering the high accuracy of Trefftz approximations.
One such mechanism is trigonometric interpolation, which itself is known
to be surprisingly accurate for smooth periodic functions, 
in comparison with other typical forms of interpolation.
We also outline, on physical grounds, a curious connection of Trefftz approximations
with the theory of random matrices.

It is hoped that these considerations will stimulate further mathematical research
and practical applications of Trefftz-based methods.

\section*{Acknowledgment}

The work of IT was supported in part by the US National Science Foundation Grants
DMS-1216927 and DMS-1620112.
The research of SM and YC was supported by the Singapore MOE Academic Research
Fund Tier 2 Grant MOE2016-T2-1-128, the Singapore MOE Academic
Research Fund Tier 2 Grant MOE2015-T2-2-008, and the Singapore MOE
Academic Research Fund Tier 3 Grant MOE2016-T3-1-006. 
The work of VM was supported in part by the US National Science Foundation Grants
DMS-1216970.

IT thanks Ralf Hiptmair, Andrea Moiola, Lise-Marie Imbert-G\'{e}rard
and Ben Schweizer for discussions.

%
%
%
%


\section*{References}

\bibliography{Igor_reference_dbase,flame-slab-references}

\end{document}